\documentclass[sigconf,anonymous=false]{acmart}

\AtBeginDocument{%
  }
\usepackage{xspace}
\usepackage{graphicx}   
\usepackage{pdfpages}
\usepackage{natbib}
\usepackage{multirow}
\usepackage{gensymb}
\setcopyright{acmlicensed}
\copyrightyear{2018}
\acmYear{2018}
\acmDOI{XXXXXXX.XXXXXXX}
\acmConference[Conference acronym 'XX]{Make sure to enter the correct
  conference title from your rights confirmation email}{June 03--05,
  2018}{Woodstock, NY}
\acmISBN{978-1-4503-XXXX-X/2018/06}

\newcommand{\ourmethod}{\textit{VibraWave}\xspace}

\begin{document}

\title{\ourmethod: Sensing the Pulse of Polluted Waters}

\author{Sagnik Ghosh}
\email{sgsagnikghosh@gmail.com}
\affiliation{%
  \institution{Indian Institute of Technology}
  \city{Kharagpur}
  \state{WestBengal}
  \country{India}
}

\author{Sandip Chakraborty}
\email{sandipchkraborty@gmail.com}
\affiliation{%
  \institution{Indian Institute of Technology}
  \city{Kharagpur}
  \state{WestBengal}
  \country{India}}

\renewcommand{\shortauthors}{Ghosh et al.}

\begin{abstract}
Conventional methods for water pollutant detection, such as chemical assays and optical spectroscopy, are often invasive, expensive, and unsuitable for real-time, portable monitoring. In this paper, we introduce \ourmethod, a novel non-invasive sensing framework that combines mmWave radar with controlled acoustic excitation, tensor decomposition, and deep learning to detect and quantify a wide range of water pollutants. By capturing radar reflections as a three-dimensional tensor encoding phase dynamics, range bin power, and angle-of-arrival (AoA), we apply PARAFAC decomposition with non-negative constraints to extract compact, interpretable pollutant fingerprints. These are used to train a lightweight student neural network via knowledge distillation, enabling joint classification and quantification of heavy metals (Cu, Fe, Mg), oil emulsions, and sediments. Extensive experiments show that \ourmethod achieves high accuracy and low RMSE across pure, binary, and tertiary mixtures, while remaining robust and computationally efficient, making it well-suited for scalable, real-time water quality monitoring.
\end{abstract}




\maketitle
\section{Introduction}
Water pollution remains a critical global challenge, threatening both environmental ecosystems and human health. The growing presence of hazardous contaminants, including heavy metals, organic residues, and suspended solids, in surface and drinking water has raised serious concerns, necessitating accurate, timely, and scalable monitoring solutions. Traditional water quality monitoring techniques, such as laboratory chemical assays, mass spectrometry, and spectroscopic analysis, are highly accurate but require extensive sample preparation, costly equipment, and skilled personnel. These approaches are fundamentally limited in their ability to support rapid, in-situ, and continuous monitoring at scale~\cite{shtull2025insights}.

Recent advances have explored portable and IoT-integrated water sensing platforms~\cite{rana2025real, georgantas2025integrated, gulifardo2025ipond,syafrudin2025review}. However, these solutions still depend on invasive probes that demand frequent calibration and are prone to degradation in harsh or remote environments~\cite{oppermann2024low}. Meanwhile, remote sensing techniques~\cite{chen2025application} provide excellent spatial coverage but fall short in resolution, particularly when it comes to detecting and quantifying specific pollutant types such as copper, oil emulsions, or sediment particulates. Machine learning frameworks built atop these platforms~\cite{liang2025monitoring, irwan2025river,baena2025intelligent,frincu2025artificial} suffer from the fundamental bottleneck of sensor-level data quality, often capturing only low-dimensional or aggregated pollutant indicators~\cite{song2025integrated,nagothu2025advancing}. Hence, despite significant progress, current technologies do not address the problem of accurate, non-invasive, multi-component pollutant identification and quantification in realistic field scenarios.

Millimeter-wave (mmWave) radar presents a compelling alternative~\cite{prabhakara2023radarhd,zheng2024enhancing,qian2022millimirror,chae2024mmcomb,soltanaghaei2021millimetro,mehrotra2024hydra}. It offers high-resolution, contactless sensing capabilities, is immune to electromagnetic interference from water, and is readily integrable into portable and embedded systems~\cite{vemuri2024deploying,guan2023neural,guan2020through,kamari2023mmsv}. In recent years, mmWave radar has shown promise in human activity detection~\cite{ji2022sifall,kamari2023environment}, gesture recognition~\cite{ma2025mmet},privacy monitoring~\cite{hu2022milliear,mei2024mmspyvr}, and even some material classification tasks~\cite{salami2023water, cao2022mmliquid}. However, its potential for water pollution sensing remains largely unexplored. Existing mmWave implementations in this domain either depend on static reflectors~\cite{cao2022mmliquid}, require heavily tuned deep models on raw IQ data~\cite{salami2023water}, or suffer from poor generalization across container types~\cite{wang2024liqdetector}.

The core insight of our work is that low-frequency acoustic excitation, when coupled with mmWave radar, induces measurable, pollutant-specific perturbations in phase, power, and spatial scattering. These modulations, distributed across frequency, space, and angle, encode rich latent structure captured by a three-dimensional radar response tensor. To leverage this structure, we propose \textbf{\ourmethod}, a physics-guided sensing pipeline integrating acoustic-induced mmWave excitation, tensor decomposition, unmixing, and neural inference for robust, non-invasive pollutant monitoring.

\noindent \textbf{Research Challenges:} Despite the potential of mmWave radar, leveraging it for multi-pollutant aqueous sensing presents unique technical hurdles. These challenges are twofold:
\begin{itemize}
    \item \textbf{C1: Extracting component-specific features from multi-dimensional radar signals.} Unlike traditional applications, the radar return from mixed chemical solutions lacks strong, discrete reflectors. Capturing subtle pollutant-induced effects across frequency, range, and AoA dimensions requires novel signal processing.
    \item \textbf{C2: Interpreting and unmixing overlapping pollutant signatures.} In real-world scenarios, water samples often contain multiple co-existing pollutants. Disentangling their individual contributions and estimating their concentrations is a challenging inverse problem that requires appropriate modeling assumptions or structured decomposition.
\end{itemize}

\noindent \textbf{Our Contributions:} To address these challenges, we develop a comprehensive radar-based sensing and modeling pipeline tailored to the unique complexities of aqueous pollutant detection. Our approach integrates physics-driven signal processing, interpretable decomposition, and efficient machine learning to deliver both accuracy and real-world applicability. Our contributions are as follows. 
\begin{itemize}
    \item \textbf{A novel active mmWave sensing framework for water pollutants.} We propose a new sensing modality that combines 77~GHz radar with low-frequency acoustic excitation to induce surface vibrations in water. These vibrations amplify pollutant-specific interactions with the radar signal, modulating phase and angle-of-arrival (AoA) based on each contaminant’s mechanical and dielectric properties. By modeling electromagnetic reflections from containerized water, and analyzing them via Capon beamforming and tone-wise processing, we extract informative latent factors and cross-dimensional correlations, enabling robust detection and quantification than conventional methods.
    
    \item \textbf{Tensor decomposition for component fingerprinting.} We employ rank-$k$ PARAFAC decomposition over the 3D radar tensors (phase $\times$ range bin $\times$ AoA), allowing extraction of interpretable, component-specific latent vectors. The method generalizes to mixtures by leveraging multilinear structure and provides direct mapping between signatures and physical pollutants.

    \item \textbf{Physics-informed NNLS unmixing.} Using the PARAFAC-derived dictionary, we solve a non-negative least squares (NNLS) problem to recover pollutant mixing ratios. This interpretable step enforces convexity and consistency with the additive electromagnetic phase response, yielding a grounded estimate of relative concentration.

    \item \textbf{Knowledge-distilled ResMLP for joint classification and regression.} We develop a compact, lightweight neural model that learns both the classification of present pollutants and their mixture proportions. A Random Forest teacher guides the network via distillation. The resulting ResMLP achieves high performance with reduced training overhead and improved generalization.

    \item \textbf{Extensive real-world evaluation across pollutant types and mixtures.} We validate our system on a diverse, manually constructed dataset of Cu$^{2+}$, Fe$^{2+}$, Mg$^{2+}$, oil, and sediment, including pure, binary, and ternary combinations\footnote{The codebase is available at \url{https://anonymous.4open.science/r/VibraWave-0D18/} (Anonymized for double blind review). All links in this paper have been last accessed on \today.}. All concentrations are prepared based on WHO drinking water standards~\cite{cpcb} to ensure practical relevance. Our pipeline achieves an overall classification accuracy of 0.85 and per-component regression RMSE of 0.20 -- well within the range achieved by lab-grade LC-HRMS (0.09--0.25~\cite{sepman2023bypassing}). It remains robust under radar tilt ($\pm$15°), acoustic speaker strength, and reflector type, underscoring its practical stability.
\end{itemize}
In summary, \ourmethod introduces a novel approach to environmental monitoring by combining high-frequency radar physics, tensor factorization, and distilled learning. It offers an end-to-end solution for real-time, non-invasive, multi-pollutant water sensing, bridging a critical gap between modern sensing and urgent environmental demands.

\section{Related Work}
Water quality sensing has traditionally relied on a mix of invasive and non-invasive approaches, each balancing sensitivity, portability, and operational complexity~\cite{kumar2024situ}. Recent developments span chemical assays, electrochemical probes, optical techniques, and electromagnetic sensing. 

\noindent\textbf{Invasive Chemical Assays:} Traditional lab-based methods such as titration, Ultraviolet–Visible (UV–Vis) spectrophotometry, atomic absorption, and ion chromatography offer high selectivity and sensitivity, often detecting contaminants at parts-per-billion levels. However, they require infrastructure, skilled personnel, and reagents, making them impractical for real-time field monitoring~\cite{sohrabi2021recent}. As Liang et al.~\cite{liang2021fg} note, these methods provide snapshot data, not continuous measurements. Their limitations in latency, cost, and scalability have prompted interest in portable alternatives.

\noindent\textbf{Electrochemical Sensors:} Ion-selective electrodes and electrochemical biosensors are attractive for in-situ monitoring of ionic and organic analytes~\cite{sulthana2024electrochemical,hui2022recent}. These systems are low-cost, disposable, and compatible with wireless platforms. However, they often suffer from fouling, signal drift, and cross-sensitivity. Additionally, they typically support single-analyte detection and require regular calibration and maintenance, limiting robustness in complex, real-world water environments.

\noindent\textbf{Optical Sensing:} UV–Vis absorption, fluorescence, and light scattering methods support multi-parameter sensing for turbidity, dissolved organics, nitrates, and microbial content~\cite{herrera2023optical,shi2022applications}. Field-deployable spectrophotometers and fluorometers can offer real-time monitoring with compact footprints. However, their performance degrades in turbid or biofouled samples, and they require careful calibration due to temperature or pH sensitivity.

\noindent\textbf{Radar and RF-Based Techniques:} Radar sensing, especially in the millimeter-wave (mmWave) range, has emerged as a non-contact method to analyze liquid properties by exploiting variations in dielectric response~\cite{skaria2022machine,liu2024microwave,joseph2025impact}. mmWave radars offer high spatial resolution and are easily miniaturized~\cite{cardamis2025leafeon}. Recent efforts include wearable radar systems for water classification~\cite{salami2023water} and static setups for liquid identification~\cite{cao2022mmliquid}. However, many systems directly process raw I/Q signals, which include noise and require handcrafted filtering. The method in~\cite{cao2022mmliquid} also depends on fixed reflective steel plates—an impractical assumption for real-world deployment. The \textit{LiqDetector} system~\cite{wang2024liqdetector} attempts container-independent sensing but remains sensitive to placement angle and material, limiting generalization across unseen containers or new liquids.

Moreover, radar-based approaches face physical constraints such as limited penetration depth, signal attenuation in liquids, and sensitivity to environmental clutter. The non-linear mapping between radar features (e.g., phase shift, reflection strength) and pollutant content necessitates data-driven models to extract meaningful patterns. To the best of our knowledge, this is the first work that performs non-invasive water pollution quantification using mmWave radar, going beyond prior efforts focused solely on classification of liquid types. Unlike optical or electrochemical sensors, which require direct contact or transparent media, our radar-based system operates non-invasively and is resilient to turbidity, making it better suited for opaque or particulate-rich samples.

\section{Preliminaries and Core Idea}
We begin by outlining the foundational principles of mmWave radar sensing, followed by a discussion of the core conceptual framework of our system.

\subsection{Basics of mmWave Sensing}
Millimeter-wave (mmWave) Frequency-Modulated Continuous Wave (FMCW) radar systems function by transmitting chirped waveforms, signals in which the frequency increases linearly with time, across a predefined bandwidth. These waveforms, when incident upon reflective targets such as suspended particulates or interfaces within a medium, produce echoes that are received by the radar. By mixing the transmitted and received signals, a process known as \textit{dechirping}, the system generates an intermediate frequency (IF) signal. This IF signal encapsulates key information about the location and movement of targets.

To extract range information, the system determines the beat frequency $f_b$, defined as the frequency difference between the transmitted and received chirps. If $T_c$ is the duration of each chirp and $B$ is the total bandwidth, then the frequency slope $S$ is given by $S = \frac{B}{T_c}$. The distance to the reflecting target ($R$) is:
\[
R = \frac{c \cdot f_b}{2S}
\]
where $c$ is the speed of light. A Range Fast Fourier Transform (Range-FFT) applied to the IF signal yields a one-dimensional range profile, with each bin corresponding to the reflected signal power at a specific distance.

To estimate target velocity, the radar transmits $N$ chirps sequentially, separated by a time interval $T_r$. A moving target introduces a phase shift between the returns of successive chirps. The phase difference $\Delta\phi$ due to a velocity $v$ is:
\[
\Delta\phi = \frac{4\pi v T_r}{\lambda}
\]
where $\lambda$ is the wavelength of the radar signal. Applying a second Fourier Transform across the $N$ chirps results in the Doppler-FFT, producing a two-dimensional range-Doppler heatmap that captures both distance and velocity information of the targets.

\subsection{Core Idea}
\label{sec:core_idea}
In the context of aqueous pollutant detection, the radar’s capabilities are extended beyond classical range and motion estimation. We introduce controlled acoustic excitations, specifically, frequency sweeps in the range of 25–125 Hz, into the liquid sample using an external actuator. These excitations induce periodic vibrations at the air-liquid interface, modulating its reflective properties based on the mechanical and electromagnetic characteristics of the solutes.

The mmWave radar, positioned above the container, observes the backscattered signal variations that result from these acoustically driven surface perturbations. Because the physical properties of different pollutants (e.g., density, viscosity, permittivity) influence the fluid's acoustic response and its electromagnetic reflection differently, these modulations serve as unique fingerprints.

The radar collects data across multiple acoustic tones, range bins, and angles of arrival (AoA), producing a complex-valued three-dimensional tensor. This tensor captures:
(1) phase variations per acoustic tone,
(2) reflected signal intensity across selected range bins, and
(3) directional response from different AoAs.

\begin{figure}[ht]
\centering
\includegraphics[width=0.45\textwidth]{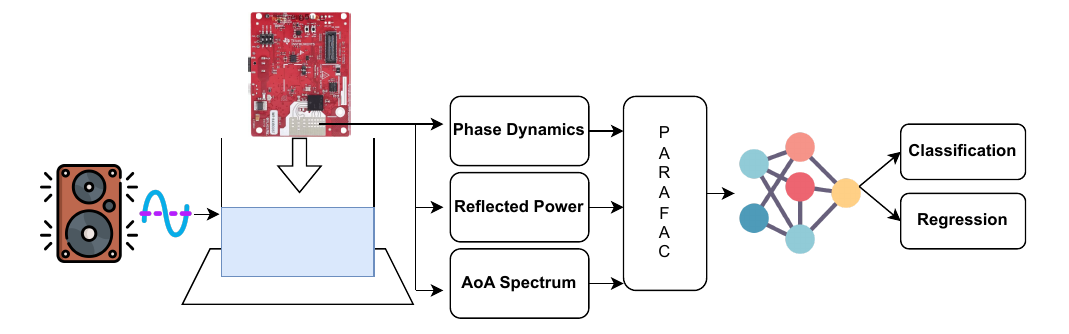}
\caption{Block diagram of \ourmethod.}
\Description{Block Diagram.}
\label{fig:block_diagram}
\end{figure}

As shown in Figure~\ref{fig:block_diagram}, the liquid sample is placed in a container situated on a reflective surface (which may be the container base itself). An acoustic transducer, such as a speaker, generates a frequency sweep that excites the fluid. The radar mounted above captures the resulting modulations in reflected signals.

These modulations are inherently pollutant-specific. They arise from how contaminants influence fluid impedance, damping, and surface resonance. By organizing the radar returns into a tensor indexed by tone frequency, range, and AoA, we retain the multidimensional nature of the interactions. This structure forms the basis for subsequent tensor decomposition and machine learning models that enable precise detection and quantification of pollutants.

\section{Electromagnetic Reflections in Vibrated Liquids}
Building upon the principles of mmWave radar sensing and the pollutant-induced surface dynamics discussed in Section~\ref{sec:core_idea}, we now formalize how these dynamic interfaces influence radar reflections. In particular, we analyze the variation in reflected signal strength as a function of dielectric impedance mismatch, interference from internal reflections, and phase modulation caused by acoustic vibrations.

\subsection{Multilayer Dielectric Model for Signal Reflection}
\label{sec:reflections}
\begin{figure}[!t]
\centering
\includegraphics[width=0.20\textwidth]{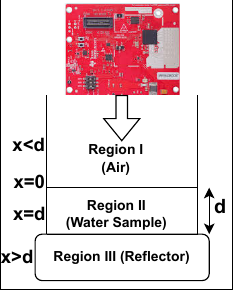}
\caption{Electromagnetic wave interaction at an air–liquid–substrate interface.}
\Description{Diagram showing the calculation of reflected signal strengths across three regions.}
\label{fig:image_label1}
\end{figure}

Consider a normal-incidence plane wave impinging upon a slab of finite thickness (e.g., a pollutant-laden liquid column) sandwiched between air and a reflective surface, as shown in Figure~\ref{fig:image_label1}. Multiple reflections and transmissions arise due to impedance mismatches at the interfaces. We now derive the net reflection coefficient \( R \) and transmission coefficient \( T \) for such a three-layer dielectric system~\cite{balanis2024balanis}.
We define the system as:
\begin{itemize}
  \item \textbf{Region I (\( x < 0 \))}: Incident medium (air) with impedance \( Z_1 \), wavenumber (spatial frequency of a wave -- that is, how many wave cycles occur per unit of distance) \( k_1 = \omega\sqrt{\mu_0\varepsilon_1} \).
  \item \textbf{Region II (\( 0 \le x \le d \))}: Water sample with impedance \( Z_2 \), wavenumber \( k_2 = \omega\sqrt{\mu_0\varepsilon_2} \), and thickness \( d \).
  \item \textbf{Region III (\( x > d \))}: Transmitting or reflecting medium (e.g., air or a solid reflector), impedance \( Z_3 \), wavenumber \( k_3 \). Often, \( Z_3 = Z_1 \), when we have air on both sides.
\end{itemize}
Here, $\omega$ is the angular frequency of the wave, $\mu_0$ is the permeability of free space (a physical constant), and $\varepsilon_1$ \& $\varepsilon_2$ are the permittivity of the two mediums, air and water, respectively.

Notably, the intrinsic impedance ($Z$) of a medium is given by the following equation~\cite{balanis2024balanis}.
\begin{equation}
Z = \sqrt{\frac{\mu}{\varepsilon}}, \quad \text{and for non-magnetic media:} \quad Z = \frac{Z_0}{\sqrt{\varepsilon_r}}, \quad Z_0 \approx 377\,\Omega.
\end{equation}
where $\mu$ represents the magnetic permeability of the medium. 

Similarly, at an interface between two media with impedances \( Z_a \) and \( Z_b \), the Fresnel coefficients $r_{ab}$ and $t_{ab}$ for normal incidence~\cite{balanis2024balanis} are given as:
\begin{equation}
r_{ab} = \frac{Z_b - Z_a}{Z_b + Z_a}, \quad t_{ab} = \frac{2Z_b}{Z_a + Z_b}.
\end{equation}
The Fresnel coefficients quantify how electromagnetic waves reflect and transmit at the interface between two media with different impedances. These coefficients are critical in modeling multilayer interactions, as they govern the amplitude and phase of the reflected radar signal based on material properties. By incorporating Fresnel reflections into our model, we can analytically predict how pollutant-induced changes in permittivity influence the radar return, thereby enabling material differentiation and concentration estimation.

\subsubsection{Analytical Derivation of the Reflection Coefficient}
To analyze how the electromagnetic wave propagates through and reflects from the multilayer structure, we express the electric field in each region as a superposition of forward- and backward-propagating plane waves. Each wave is represented as a complex exponential whose phase evolves according to the wavenumber \( k \) in that region. The fields are continuous at the boundaries, and their behavior is governed by Maxwell's equations~\cite{balanis2024balanis}. The electric field distributions of the electromagnetic wave in different spatial regions ($E_I(x)$, $E_{II}(x)$, and $E_{III}(x)$) of the multilayer medium is expressed as follows. $E_i$ and $E_r$ are incident and reflected electric fields. $E_t$ is the transmitted electric field.
\begin{align*}
\text{Region I } (x<0):\quad & E_I(x) = E_i\, e^{-jk_1 x} + E_r\, e^{jk_1 x}, \\
\text{Region II } (0 \le x \le d):\quad & E_{II}(x) = A\, e^{-jk_2 x} + B\, e^{jk_2 x}, \\
\text{Region III } (x>d):\quad & E_{III}(x) = E_t\, e^{-jk_3(x-d)}.
\end{align*}
The total reflected field in Region I arises from a combination of direct reflections at the air–liquid interface and multiple internal reflections within the liquid slab. These reflections can be quantified using the field expressions derived above, with each term corresponding to a physical interaction at an interface. Accordingly, we compute the individual contributions to the total reflected field arising from each interface interaction as follows.

\paragraph{Reflection Contributions:}
At \( x = 0 \), a portion \( r_{12} \) is reflected directly:
\[
E_r^{(0)} = E_i \, r_{12}.
\]

The transmitted part undergoes internal reflections inside the slab. The first internal round trip contributes:
\[
E_r^{(1)} = E_i \, t_{12} \, t_{23} \, r_{23} \, e^{-j2k_2 d}.
\]

Subsequent internal reflections contribute an infinite geometric series due to repeated round trips:
\[
E_r = E_i \, r_{12} + E_i\, t_{12} \, t_{23} \, r_{23} \, e^{-j2k_2 d} \sum_{n=0}^{\infty} (r_{21} \, r_{23} \, e^{-j2k_2 d})^n.
\]

The closed-form expression for the net reflection coefficient is:
\[
R = \frac{E_r}{E_i} = r_{12} + \frac{t_{12} \, t_{23} \, r_{23} \, e^{-j2k_2 d}}{1 - r_{21} \, r_{23} \, e^{-j2k_2 d}}.
\]

\paragraph{Special Case: Symmetric Environment (\( Z_1 = Z_3 \))}

If the medium on either side of the liquid slab is the same (e.g., air), then:
\[
r_{23} = r_{21}, \quad \text{and} \quad r_{21} = -r_{12}.
\]

This simplifies the reflection coefficient to:
\begin{equation}
R = r_{12} + \frac{t_{12} \, t_{23} \, (-r_{12}) \, e^{-j2k_2 d}}{1 - r_{12}^2 \, e^{-j2k_2 d}}.
\end{equation}

\subsubsection{Physical Interpretation: Signal Strength and Loss}
Since the radar receives backscattered power, the received signal strength is proportional to:
\(
P_{\text{reflected}} \propto |R|^2 \cdot P_i
\),
where $P_i$ is the incident power.
In lossy media (e.g., polluted water), the wavenumber becomes complex:
\begin{equation}
    k_2' = k_2 - j\alpha,
\end{equation}
where \( \alpha \) is the attenuation constant. This introduces exponential decay in the transmitted and reflected waves, attenuating the signal. Hence, both impedance mismatch and internal absorption influence the amplitude of the radar returns.

\subsection{Phase Shift Induced by Acoustic Vibration}
\label{sec:phase_vibration}
In the previous sub-section, we modeled amplitude variations via multilayer dielectric theory. Here, we describe phase changes caused by sub-wavelength acoustic perturbations of the liquid-air interface.

The transmitted FMCW chirp reflects from a vibrating surface at range \( R \) with a round-trip delay:
\(
\tau = \frac{2R}{c},
\) where $c$ is the speed of light. This induces a beat frequency \( f_b \), and the IF signal becomes:
\begin{equation}
s_{\text{IF}}(t) = A \sin(2\pi f_b t + \phi_0).
\end{equation}

Its complex baseband representation is:
\begin{equation}
    s[n] = A e^{j(2\pi f_b n T_c + \phi[n])}.
\end{equation}

If the interface vibrates by a displacement \( \Delta d[n] \ll \lambda \), the phase changes as:
\begin{equation}
    \phi[n] = \phi_0 + \frac{4\pi}{\lambda} \Delta d[n],
\end{equation}
and displacement can be recovered via:
\begin{equation}
    \Delta d[n] = \frac{\lambda}{4\pi} (\phi[n] - \phi_0).
\end{equation}
This phase modulation arises because acoustic vibrations at the liquid-air interface induce tiny vertical displacements -- typically much smaller than the radar wavelength (\( \Delta d[n] \ll \lambda \)). Although these displacements are sub-wavelength in scale, they produce measurable shifts in the phase of the radar’s intermediate-frequency (IF) signal due to the round-trip nature of wave propagation. Since different pollutants alter the mechanical and dielectric properties of the fluid (e.g., by changing viscosity, density, or permittivity), they modulate the acoustic response of the interface differently. As a result, the phase sequence \( \phi[n] \) encodes pollutant-specific vibrational signatures. 

By precisely tracking these phase variations across successive chirps, \ourmethod\ is able to non-invasively sense pollutant-induced dynamics through the radar signal, enabling rich, time-resolved characterization of fluid properties at the surface.

\subsection{Pilot Experimental Validation}
To validate the preceding theoretical framework, we conducted pilot studies investigating the effects of material composition and concentration on radar signal characteristics.

\subsubsection{Power-Based Material Differentiation}
To empirically validate the theoretical predictions from our multilayer reflection model, we conducted a series of controlled experiments using distilled water samples mixed with various pollutants at concentrations ranging from 2–5 mg/L. These pollutants were selected to represent a range of chemical properties, including ionic content, organic compounds, and suspended solids, each of which influences the liquid's dielectric behavior. According to the multilayer dielectric model described earlier, the reflected power depends on both the intrinsic impedance mismatch between the liquid and air (\( Z_2 \) vs. \( Z_1 \)) and the attenuation within the slab, governed by its complex wavenumber \( k_2' = k_2 - j\alpha \). Since each contaminant alters the water's effective permittivity \( \varepsilon_2 \) and attenuation constant \( \alpha \), the reflection coefficient \( R \), and consequently the received power \( |R|^2 \), varies accordingly.

\begin{figure}[!t]
\centering
\begin{minipage}[t]{0.48\linewidth}
    \centering
    \includegraphics[width=\linewidth]{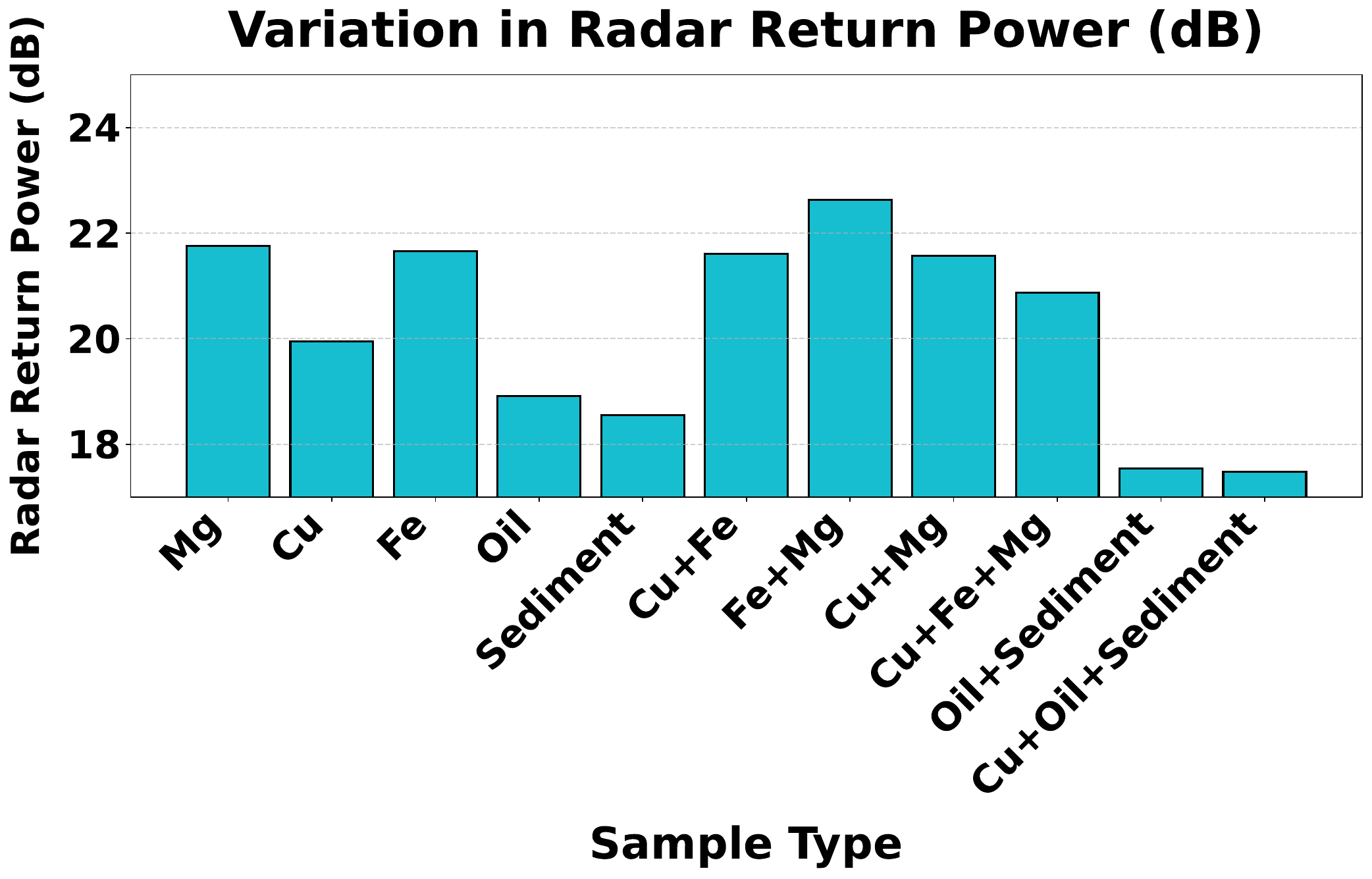}
    \caption*{(a) Radar power variations}
\end{minipage}
\hfill
\begin{minipage}[t]{0.48\linewidth}
    \centering
    \includegraphics[width=\linewidth]{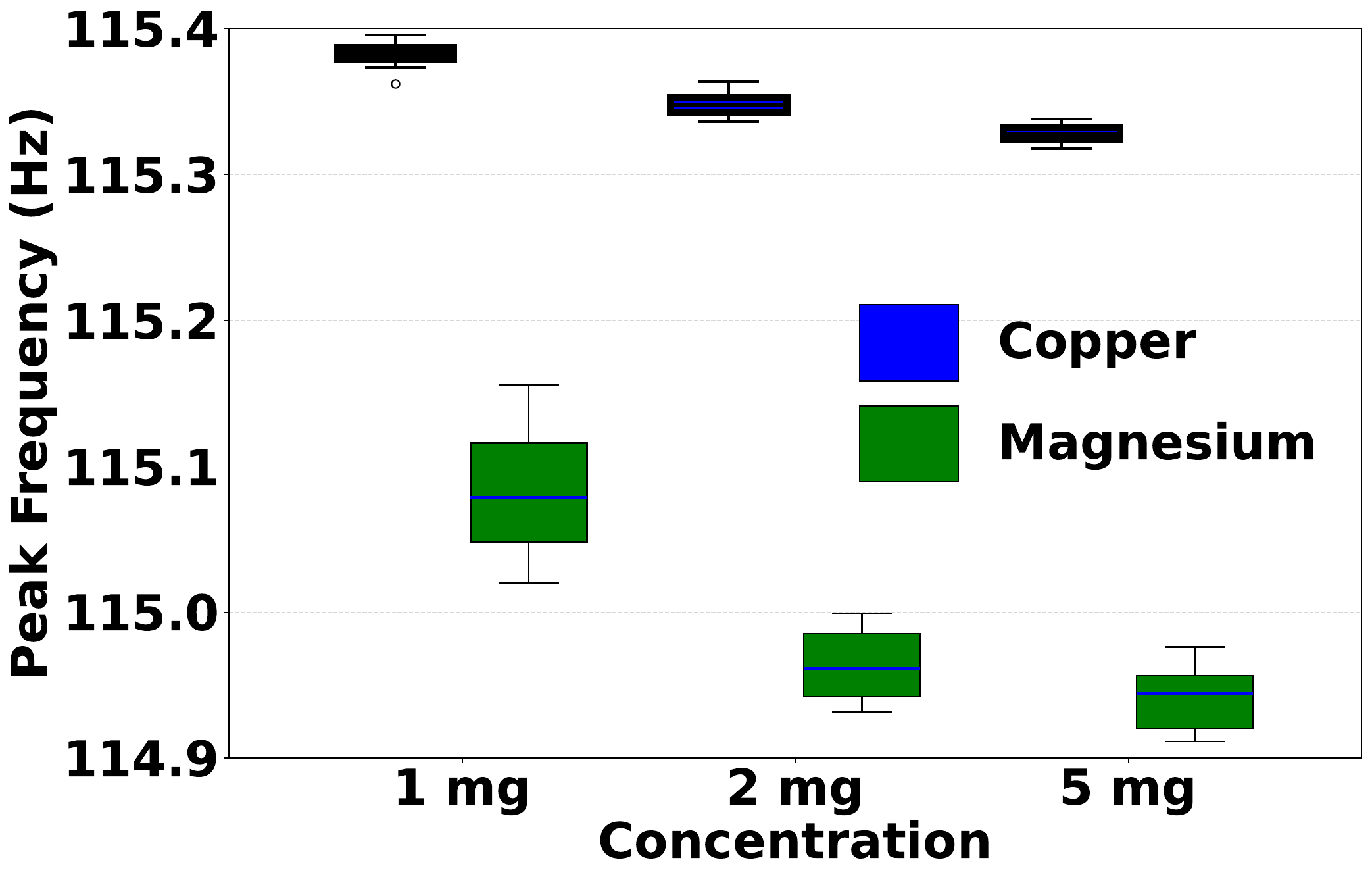}
    \caption*{(b) Peak stability across different concentrations of copper and magnesium}
\end{minipage}
\caption{Variations in reflected power and peak frequency stability}
\Description{Variations in reflected power and peak frequency stability.}
\label{fig:sidebyside_diagrams2}
\vspace{-0.5em}
\end{figure}

As shown in Figure~\ref{fig:sidebyside_diagrams2}(a), these variations are evident in the measured radar return power. The radar reliably distinguishes between the pollutant-contaminated samples based on their dielectric response, supporting the hypothesis that solute-induced changes in electromagnetic impedance directly influence reflected signal strength.

\subsubsection{Phase and Frequency Response to Concentration Variations}
To further assess the robustness of our sensing methodology, we examined how phase and frequency responses vary with changes in pollutant concentration. Specifically, we conducted a series of measurements using distilled water mixed with varying concentrations of copper (Cu) and magnesium (Mg) ions. Each sample was prepared within a range of 2–5\,mg/L, and measurements were recorded using the same mmWave radar setup and acoustic excitation protocol. We analyzed the radar's IF phase traces and extracted the dominant frequency components induced by the sinusoidal vibration. According to the phase model discussed previously, pollutant-induced surface vibrations modulate the radar phase response via sub-wavelength displacements, yielding a phase shift of the form \( \phi[n] = \phi_0 + \frac{4\pi}{\lambda} \Delta d[n] \). The tone-specific frequency response stems from the mechanical resonance characteristics of the liquid sample under acoustic excitation. As pollutant concentration increases, changes in the fluid’s damping or compliance can manifest as shifts in the peak response frequency.

As shown in Figure~\ref{fig:sidebyside_diagrams2}(b), we observe a subtle but consistent drop in the peak frequency as pollutant concentration increases -- validating the expected behavior from our multilayer resonance model. While the frequency shift is not very high, it is sufficiently measurable (statistically significant, $p<0.05$) and opens the door to learning concentration-dependent variations. This suggests that, once the pollutant type is identified, a regression model can be trained on peak frequency features to estimate its concentration accurately.

\subsubsection{Selection of Acoustic Excitation Profile}
We compared three standard excitation schemes: sinusoidal, random, and shock vibrations, and found that sinusoidal frequency sweeps (25–125 Hz) yield clean, repeatable resonance-induced modulations in the radar signal. This reinforces the design choice in Figure~\ref{fig:block_diagram}, where a controlled sinusoidal speaker excites the water sample for pollutant-specific response extraction~\cite{shanbhag2023contactless}.

\section{Methodology}
\label{sec:methodology}
\begin{figure}[ht]
\centering
\includegraphics[width=0.47\textwidth]{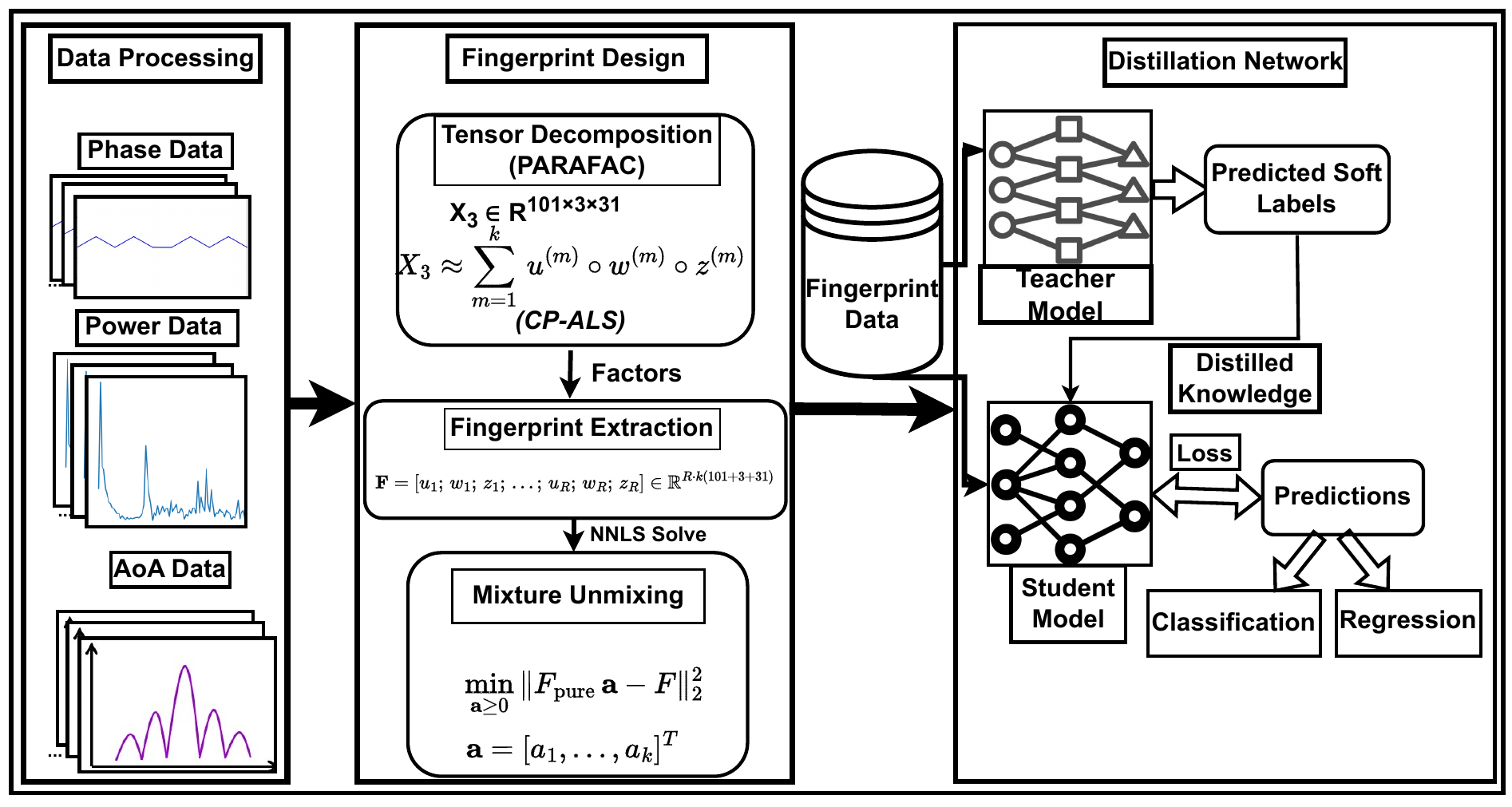}
\caption{System block diagram for end-to-end pollutant detection using \ourmethod.}
\Description{Diagram showing the system block diagram.}
\label{fig:block_diagram2}
\end{figure}

Building on our electromagnetic modeling of liquid interfaces under acoustic excitation and subsequent empirical pilot studies, we now describe the end-to-end system architecture for detecting and quantifying aqueous pollutants using mmWave radar. The overall pipeline comprises four stages: (1) radar data acquisition during acoustic excitation, (2) tensor representation of phase–spatial signatures, (3) factorized fingerprint extraction and unmixing, and (4) supervised distillation using a compact neural model. Figure~\ref{fig:block_diagram2} illustrates the overall flow of the system.

\subsection{Data Acquisition and Preprocessing}
To excite pollutant-specific mechanical resonances, we use a speaker to emit sinusoidal tones sweeping from 25\,Hz to 125\,Hz at 1\,Hz resolution. Each tone is held for one second, inducing surface vibrations on the aqueous sample. Concurrently, a IWR1843BOOST mmWave radar from Texas Instruments (TI) collects I/Q ADC data at 10\,fps, yielding 1010 chirp frames per complete sweep. This frequency-controlled excitation leverages the physical mechanisms discussed in Section~\ref{sec:reflections}, where vibration-induced surface perturbations alter both the amplitude and phase of reflected mmWave signals. From each radar frame, we extract three key quantities: unwrapped phase, reflected power across a small set of range bins, and angle-of-arrival (AoA) spatial responses. These features serve as the core input dimensions for subsequent tensor modeling.

\subsection{3D Tensor Construction}
The construction of the 3D tensor representation is directly motivated by the electromagnetic modeling and signal behavior described in earlier sections. In Section~\ref{sec:reflections}, we analytically modeled how changes in the dielectric properties and thickness of the liquid layer affect the reflected signal strength due to multilayer impedance mismatches. This explains why different substances, and even varying concentrations of the same substance, produce distinct radar power profiles. In parallel, Section~\ref{sec:phase_vibration} described how surface vibrations, induced by acoustic excitation, lead to sub-wavelength displacements at the liquid–air interface. These perturbations are encoded in the radar’s phase response, allowing us to extract fine-grained signatures tied to the mechanical and dielectric properties of the fluid.

Building on these principles, we capture the multidimensional radar response -- across acoustic excitation frequency (tone), spatial range (depth), and angle of arrival -- by constructing a real-valued 3D tensor. The acoustic sweep consists of 101 discrete sinusoidal tones from 25\,Hz to 125\,Hz, with each tone held constant for one second. During this time, the mmWave radar captures data at 10 frames per second, resulting in 1010 total radar frames per sweep.

For each tone, we aggregate information across 10 frames and focus our attention on three specific range bins, \( b \in \{1, 2, 3\} \), selected based on consistent empirical observation. These bins correspond to the spatial location of the liquid surface and are identified via range-Doppler profiling as having the most stable and high-magnitude reflections, characteristics that reflect strong impedance mismatches and minimal multipath interference. By excluding bins influenced by container edges or ambient clutter, we isolate the signal features that are most directly shaped by the pollutant-laden liquid itself.

To extract angular (directional) characteristics, we apply Capon beamforming~\cite{openradar2019}, a high-resolution spatial filtering method, across the radar’s antenna array. Capon beamforming computes the spatial power spectrum for a dense grid of steering angles, generating 31 angle-of-arrival (AoA) samples across the azimuth range from $-30^\circ$ to $+30^\circ$. For each tone–range bin pair, this produces a spatial response vector capturing the energy arriving from different directions.
The final phase-aware tensor is therefore defined as:
\[
\mathbf{X}_3 \in \mathbb{R}^{101 \times 3 \times 31},
\]
where the dimensions respectively correspond to: (1) tone index (which maps to both time and acoustic frequency), (2) selected radar range bins capturing reflected power and phase from the liquid surface, and (3) AoA samples capturing directional variations via beamforming.


\subsection{Prototype Learning via Rank-$k$ PARAFAC}
To extract interpretable, pollutant-specific patterns from the 3D radar tensor, we employ the Canonical Polyadic (CP) decomposition, also known as PARAFAC (Parallel Factor Analysis)~\cite{bro1997parafac}. PARAFAC generalizes matrix factorization to higher-order tensors, representing a multi-dimensional array as a sum of outer products of rank-1 vectors. For a third-order tensor, this means decomposing it into three sets of latent factors, one per mode, such that the reconstructed tensor approximates the original structure. Each component in the decomposition captures a coherent latent signature across frequency (acoustic tones), spatial depth (range bins), and angular response (AoA). In our context, these latent signatures encode the unique radar-phase behaviors associated with each individual pollutant.

For each pure liquid sample, we begin by constructing its magnitude tensor $|\mathbf{X}_3^c| \in \mathbb{R}^{101 \times 3 \times 31}$, representing the phase dynamics across tone index $i$, selected range bins $r$, and spatial angles $a$. We then apply a rank-1 CP decomposition to extract a single pollutant-specific signature. The decomposition takes the form:
\begin{equation}
    |\mathbf{X}_3^c|_{i,r,a} \approx \sum_{k=1}^{1} u_i\, v_r\, w_a,
\end{equation}
with the factor matrices defined as:
\[
U \in \mathbb{R}^{101 \times 1},\quad V \in \mathbb{R}^{3 \times 1},\quad W \in \mathbb{R}^{31 \times 1},
\]
where \( u_i \), \( v_r \), and \( w_a \) respectively capture the frequency-dependent, depth-wise, and angular radar response patterns for that pollutant.

We solve this decomposition using the Alternating Least Squares (ALS) algorithm, a standard method that optimizes each factor matrix iteratively while keeping the others fixed. The updates follow:
\begin{align*}
U &\leftarrow \arg\min_{U'} \left\| |\mathbf{X}_3^c|_{(1)} - (W \odot V)(U')^\top \right\|_F^2, \\
V &\leftarrow \arg\min_{V'} \left\| |\mathbf{X}_3^c|_{(2)} - (W \odot U)(V')^\top \right\|_F^2, \\
W &\leftarrow \arg\min_{W'} \left\| |\mathbf{X}_3^c|_{(3)} - (V \odot U)(W')^\top \right\|_F^2,
\end{align*}
where \( \odot \) denotes the Khatri-Rao (column-wise Kronecker) product, and \( |\mathbf{X}_3^c|_{(n)} \) denotes the mode-$n$ matricized unfolding of the tensor.

Once the ALS algorithm converges, we normalize each factor to have unit $\ell_2$ norm and absorb the scale into a scalar weight. The final pollutant fingerprint is formed by concatenating the normalized vectors:
\[
\mathbf{p}_c = [u; v; w] \in \mathbb{R}^{135},
\]
yielding a 135-dimensional signature that compactly represents the radar response of the pure pollutant. Collecting such vectors across all $C$ pure substances constructs the dictionary:
\[
\mathbf{P} = [\mathbf{p}_1, \dots, \mathbf{p}_C] \in \mathbb{R}^{135 \times C}.
\]

For mixed samples, which may consist of two or three pollutants, we apply a rank-$k$ CP decomposition with $k = 2$ or $k = 3$ respectively. This yields $k$ factor triples \( \{ (u_k, v_k, w_k) \}_{k=1}^K \), each representing the radar signature of an individual component within the mixture. These are concatenated to form the mixture fingerprint:
\[
\mathbf{p} = [u_1;\, v_1;\, w_1;\, \dots;\, u_k;\, v_k;\, w_k] \in \mathbb{R}^{k(101 + 3 + 31)},
\]
which serves as a latent representation encoding both the presence and proportion of multiple constituents. These fingerprints form the input to downstream unmixing and classification stages.

\subsection{Non-Negative Least Squares Unmixing}
The goal of this step is to quantitatively decompose the radar-phase signature of a complex (mixed) aqueous sample into its constituent pollutant components. Building on the prototype fingerprints extracted via PARAFAC decomposition from pure samples, we now aim to determine which pollutants are present in a mixture and in what proportions, based solely on its observed tensor signature. This enables both classification and quantification of contaminants, consistent with the additive nature of radar signal responses under linear superposition.

Given a mixture fingerprint \( \mathbf{f} \in \mathbb{R}^D \) derived from a rank-$k$ PARAFAC decomposition, we express it as a convex combination of known prototype vectors by solving the following constrained least-squares problem:
\[
\hat{\mathbf{c}} = \arg\min_{\mathbf{c} \geq 0,\, \mathbf{1}^\top \mathbf{c} = 1} \left\| \mathbf{P} \mathbf{c} - \mathbf{f} \right\|_2^2,
\]
where \( \mathbf{P} \in \mathbb{R}^{D \times C} \) is the matrix of pure-component fingerprints (each column corresponding to one pollutant) and \( \hat{\mathbf{c}} \in \mathbb{R}^C \) represents the estimated mixing proportions of those components. The non-negativity constraint ensures physically meaningful (non-negative) concentrations, and the sum-to-one constraint ensures the solution is interpretable as a composition ratio.

The convex combination:
\[
\mathbf{p}_{\text{mix}} = \mathbf{P} \hat{\mathbf{c}} \in \mathbb{R}^D
\]
reconstructs the mixture fingerprint as a point in the same latent space as the pure signatures, capturing both identity and relative abundance of each component.

This unmixing step is essential for pollutant identification and is grounded in the physical observation that phase and magnitude responses in radar signals linearly superimpose under weak-scattering and small-vibration regimes. Thus, the fingerprint of a mixture naturally lies within the convex hull of the individual constituent signatures, allowing accurate reconstruction and interpretation via non-negative least squares (NNLS).

\subsection{Student–Teacher Distillation Framework}
While NNLS unmixing produces interpretable proportions, we further refine prediction using a student–teacher learning setup. The goal is to regress mixture ratios and classify component presence directly from tensor fingerprints.

\subsubsection*{Teacher Model (Random Forest)}
We train a \texttt{RandomForestRegressor} ensemble to predict mixture ratios from PARAFAC fingerprints. The output \( \hat{\mathbf{c}}^{\text{RF}} \in [0,1]^C \) serves as soft supervision for the student network.

\subsubsection*{Student Network (ResMLP)}
\begin{figure}[ht]
\centering
\includegraphics[width=0.35\textwidth]{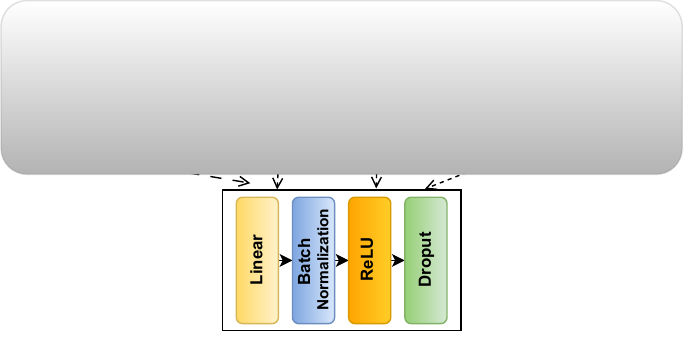}
\caption{ResMLP student network for pollutant classification and mixture regression.}
\Description{Diagram showing the ResMLP block diagram.}
\label{fig:block_diagram1}
\end{figure}

We use a compact ResMLP architecture (Figure~\ref{fig:block_diagram1}) to jointly learn regression and classification from the fingerprints:
\[
\mathbf{p} \in \mathbb{R}^{D}, \quad D = k(I + R + A).
\]

The model consists of 4 fully connected layers with batch normalization, dropout, and a residual skip connection. The output layer comprises two heads: one softmax for mixture ratio regression, and one sigmoid for presence detection. The objective function combines three losses:
\[
\mathcal{L}(\theta) = \alpha\, D_{\mathrm{KL}}\bigl(s_\theta(\mathbf p)\,\|\,\hat{\mathbf c}^{\text{RF}}\bigr) + (1 - \alpha)\sum_{j} L_\delta(s_j(\mathbf p), c_j^*) + \beta \sum_j \text{BCE}(y_j^*, \hat{y}_j),
\]
where $D_{\mathrm{KL}}$ is the distillation loss, $L_\delta$ is the Huber regression loss, and BCE is the binary cross-entropy for presence classification. We set $\alpha=0.7$, $\beta=0.5$, and use Adam optimizer with a one-cycle learning rate schedule.
The student-teacher framework enables accurate identification and quantification of pollutants by combining the interpretability of PARAFAC decomposition with the flexibility of neural network inference. While PARAFAC produces a structured factorization of the radar tensor, it requires specifying a rank \( k \), which corresponds to the assumed number of mixture components. In practice, we set \( k = n \), where \( n \) is the maximum number of possible pollutants, to allow for general mixtures. However, this overcomplete representation may include extra components that do not actually appear in the mixture. To address this, the ResMLP student model, trained using soft ratio estimates \( \hat{\mathbf{c}}^{\text{RF}} \) from the Random Forest teacher, jointly performs two tasks: (1) continuous regression of mixture concentrations and (2) binary classification of each pollutant’s presence. The classification head serves a critical role by identifying which mixture components are truly active, thereby deducing the actual number of pollutants present, even though the initial tensor decomposition assumed a larger upper bound. This allows the model to disentangle real pollutant signals from noise or decomposition artifacts in a data-driven, learnable fashion.

\section{Implementation Details}
To validate the effectiveness and generalizability of \ourmethod, we implemented a complete end-to-end system comprising mmWave radar sensing, acoustic excitation, signal processing, and deep learning-based analysis. Our evaluation spans a wide variety of common water pollutants, mixed in realistic concentrations, and carefully controlled under laboratory conditions to ensure reproducibility and accurate ground truth labeling. The following sections detail the data collection process, hardware setup, acoustic configuration, and model training protocols.

\subsection{Data Collection Protocol}
To ensure controlled and reproducible evaluation of \ourmethod, we designed a comprehensive data collection process that spans a wide diversity of pollutants, mixture types, and concentrations. All experiments were conducted under clean, lab-grade conditions to isolate and capture the true electromagnetic response of liquid pollutants. For accurate ground truth, we manually prepared each sample by mixing defined quantities of pollutants into deionized (DI) water. The concentrations were carefully chosen to reflect realistic contamination scenarios while remaining within the permissible limits specified in the WHO drinking water guidelines~\cite{cpcb}.

We constructed five distinct pollutant stock solutions:
Cu$^{2+}$ (copper sulfate), Fe$^{2+}$ (ferrous sulfate), Mg$^{2+}$ (magnesium sulphate), a light oil-in-water emulsion, and a fine sediment suspension. Each was mixed in DI water and maintained at WHO-compliant concentrations for drinking water~\cite{cpcb} at a predefined weight (0.4-5 mg/L). Pollutant concentrations were prepared using volumetric dilution with laboratory-grade stock solutions. While we do not employ ground-truth validation via chemical assay, the consistency of response trends across concentrations and pollutants supports the sensitivity and internal consistency of our radar-based measurements.

\textbf{Binary mixtures:} We considered all $\binom{5}{2} = 10$ unordered pairs of pollutants and prepared them at three volumetric ratios: 25:75, 50:50, and 75:25. Each of these $10 \times 3 = 30$ mixtures was replicated three times, yielding a total of 90 binary samples.

\textbf{Ternary mixtures:} For ternary combinations, all $\binom{5}{3} = 10$ unique pollutant triplets were used. Each triplet was mixed in four different volumetric ratios -- one symmetric (33.3:33.3:33.3) and three asymmetric (20:40:40, 40:20:40, 40:40:20), with three replicates per ratio, resulting in 120 ternary samples.

\begin{figure}[ht]
\centering
\includegraphics[width=0.4\textwidth]{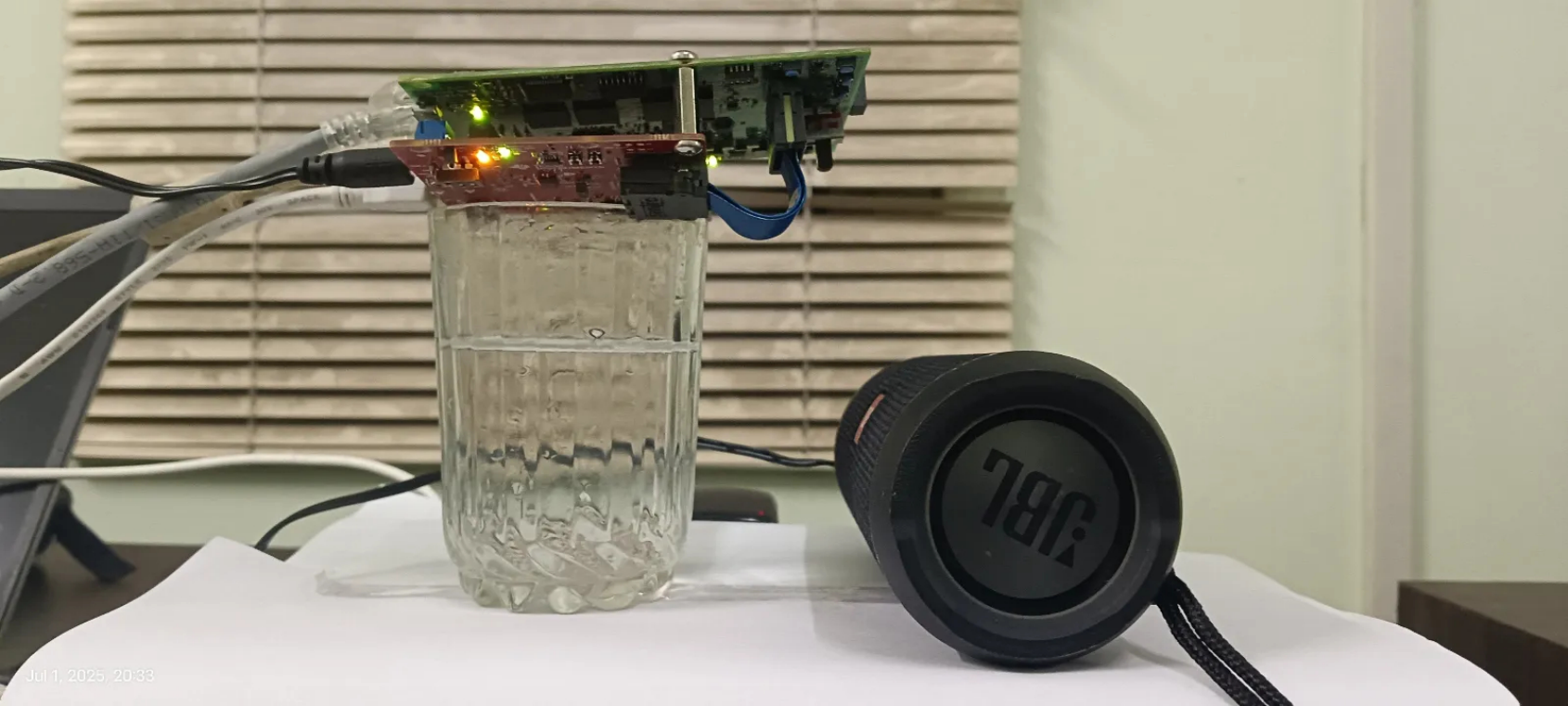}  
\caption{Data Acquisition Setup}
\Description{Block diagram of the hardware and acoustic setup used for data collection.}
\label{fig:setup}
\end{figure}

Each sample underwent a 101-second acoustic excitation sweep, with radar data captured at 10 FPS, totaling 1010 I/Q frames per sample. Samples were allowed to settle for 30 seconds after mixing before measurements. Between samples, the container was thoroughly cleaned to eliminate cross-contamination. All data was collected at a stable ambient temperature of 24°C in an acoustically isolated laboratory, with ambient noise held below 35 dBA. The overall experimental setup is shown in Figure~\ref{fig:setup}.

\subsection{Hardware Implementation}
Radar sensing was performed using the Texas Instruments (TI) IWR1843BOOST mmWave sensor, paired with a DCA1000 data capture card. The radar operates at 77 GHz and features 3 transmit and 4 receive antennas, offering high-resolution sensing capabilities. The radar configuration provided a range resolution of 4.88 cm and a Doppler resolution of 0.0806 m/s. It was mounted on a stable tripod directly above the glass container, with antennas oriented perpendicularly to the liquid surface. Data was streamed to a laptop equipped with an Intel i5 CPU (2.40 GHz), 8 GB RAM, NVIDIA GeForce MX350 GPU, and a 1TB SSD, running Windows 10.

\subsection{Software Implementation}
Radar data acquisition was managed using the OpenRadar framework~\cite{openradar2019}. We developed a custom Lua script compatible with both mmWave Studio and OpenRadar tools for seamless flashing and data capture. The raw ADC data was saved as NumPy (\texttt{.npy}) files for subsequent processing. All downstream signal processing, tensor construction, and deep learning components were implemented in Python. For benchmarking, we also ran the baseline method from~\cite{salami2023water} using the authors' publicly available code\footnote{\url{https://version.aalto.fi/gitlab/salamid1/water-quality-with-mmwave-radar}} on our dataset.

\subsection{Speaker and Acoustic Excitation}
To deliver precise and repeatable acoustic excitation, we used a JBL Flip SE 3 Bluetooth speaker placed 5~cm from the container. The excitation signal was synthesized in Python using SciPy: a linear sweep comprising 101 sinusoidal tones, ranging from 25~Hz to 125~Hz in 1~Hz steps. Each tone was 1 second long and sampled at 44.1~kHz with full 16-bit amplitude. The resulting waveform was saved as a \texttt{.wav} file and played via a smartphone at a fixed volume level, measured using a calibrated NUL-212 sound sensor to ensure consistency. Each tone's duration was synchronized with the radar's 10 FPS sampling rate to maintain alignment between acoustic and radar data streams.

\subsection{Distillation Network Training}
Both teacher and student models were implemented in PyTorch. The teacher network is a \texttt{RandomForestRegressor} ensemble, and the student network is a ResMLP architecture as described in Section~\ref{sec:methodology}. Models were trained using the Adam optimizer with a weight decay of $1 \times 10^{-6}$ and a One-Cycle learning rate schedule over 200 epochs, peaking at a learning rate of $1 \times 10^{-3}$. Gradient clipping was applied with a norm threshold of 1.0 to stabilize training. A 20\% validation split was used for early stopping. The loss terms were weighted as follows: distillation loss (KL divergence) $\alpha = 0.7$, classification loss weight $\beta = 0.5$, and Huber loss threshold $\delta = 0.1$. These hyperparameters were selected empirically to balance classification accuracy and concentration estimation.

\section{Results}
In this section, we evaluate the performance of our phase-tensor fingerprinting pipeline combined with the distilled ResMLP model on a comprehensive dataset comprising pure, binary, and tertiary water mixtures—including metal ions, oil emulsions, and sediments. We begin by assessing the model’s ability to classify the presence of individual pollutants, followed by its accuracy in regressing their mixing proportions. To test robustness, we perform evaluations across multiple system components. For both classification and regression tasks, the dataset is split into 80\% for training and 20\% for testing, stratified by mixture type. To ensure statistical reliability, we report results averaged across multiple random seeds with varying train-test splits and leave-one-out cross-validation.

\subsection{Classification Performance}
To evaluate the ability of our system to detect the presence of specific pollutants, we frame classification as a multi-label problem over five target classes: Cu$^{2+}$, Fe$^{2+}$, Mg$^{2+}$, oil, and sediment. Our phase–AoA tensor captures rich spatiotemporal signatures of each sample, and through the PARAFAC decomposition and student ResMLP pipeline, we extract discriminative fingerprints suitable for classification. We report standard metrics, accuracy, precision, recall, and F1-score, across pure, binary, and tertiary mixtures, using the model trained on 80\% of the data and tested on the remaining 20\%, with performance averaged over multiple random seeds.

\noindent\textbf{Pure Samples:}
Table~\ref{tab:classification_metrics} presents the per-class classification results for pure pollutant samples. Organic contaminants (oil and sediment) yield the highest accuracy and F1 scores (0.92/0.89 for oil, 0.90/0.86 for sediment), reflecting their strong and distinct vibrational phase–AoA patterns. In contrast, metal ions (Cu, Fe, Mg) exhibit comparatively lower accuracy (0.72–0.78) and F1-scores (0.76–0.82), due to overlapping subspace signatures among ionic species. Nonetheless, the student network achieves robust macro-averaged performance: Accuracy = 0.85, Precision = 0.85, Recall = 0.81, and F1 = 0.83. Subset accuracy (i.e., exact match across all five labels) reaches 0.80, showing reliable detection of all components in most cases.

\begin{table}[!ht]
\centering
\caption{Classification performance for pure solutions}
\scriptsize
\begin{tabular}{|c|c|c|c|c|}
\hline
\textbf{Component} & \textbf{Accuracy} & \textbf{Precision} & \textbf{Recall} & \textbf{F1-score} \\
\hline
Cu$^{2+}$    & 0.78 & 0.85 & 0.80 & 0.82 \\
Fe$^{2+}$    & 0.75 & 0.82 & 0.78 & 0.80 \\
Mg$^{2+}$    & 0.72 & 0.78 & 0.74 & 0.76 \\
Oil          & 0.92 & 0.90 & 0.88 & 0.89 \\
Sediment     & 0.90 & 0.88 & 0.85 & 0.86 \\
\hline
\end{tabular}
\label{tab:classification_metrics}
\end{table}

\noindent\textbf{Binary Mixtures:}
Classification performance on binary mixtures is shown in Table~\ref{tab:binary_mixture_metrics}. Pairs involving only metals (Cu+Fe, Cu+Mg, Fe+Mg) yield comparatively lower F1-scores (0.78–0.82), primarily due to similar phase characteristics, especially at unbalanced mixing ratios. In contrast, mixtures combining metals with oil or sediment show higher scores (F1 $\geq$ 0.85), as the unique signatures from organic components help the model disentangle overlapping patterns. The oil+sediment pair achieves the best results (Accuracy = 0.92, F1 = 0.92), highlighting their clear separability in the latent tensor space. The overall macro-averaged accuracy and F1 across binary mixtures are both 0.85, showing the model’s reliability across two-component settings.

\begin{table}[!ht]
\centering
\caption{Classification performance on binary mixtures}
\scriptsize
\begin{tabular}{|c|c|c|c|c|}
\hline
\textbf{Mixture} & \textbf{Accuracy} & \textbf{Precision} & \textbf{Recall} & \textbf{F1-score} \\
\hline
Cu + Fe          & 0.82 & 0.84 & 0.80 & 0.82 \\
Cu + Mg          & 0.80 & 0.83 & 0.78 & 0.80 \\
Cu + Oil         & 0.88 & 0.90 & 0.86 & 0.88 \\
Cu + Sediment    & 0.86 & 0.88 & 0.85 & 0.86 \\
Fe + Mg          & 0.78 & 0.80 & 0.76 & 0.78 \\
Fe + Oil         & 0.87 & 0.89 & 0.85 & 0.87 \\
Fe + Sediment    & 0.85 & 0.87 & 0.84 & 0.85 \\
Mg + Oil         & 0.86 & 0.88 & 0.85 & 0.86 \\
Mg + Sediment    & 0.84 & 0.86 & 0.83 & 0.84 \\
Oil + Sediment   & 0.92 & 0.94 & 0.91 & 0.92 \\
\hline
\end{tabular}
\label{tab:binary_mixture_metrics}
\end{table}

\noindent\textbf{Ternary Mixtures:}
Table~\ref{tab:ternary_mixture_metrics} summarizes classification results on all 10 ternary combinations. Mixtures composed solely of metals (Cu+Fe+Mg) are most difficult to classify (Accuracy = 0.70, F1 = 0.73), due to overlapping ionic signatures. Adding an organic (oil or sediment) improves separability; e.g., Cu+Fe+Oil and Cu+Fe+Sediment achieve F1 $\approx$ 0.80. Mixtures containing two organics (e.g., Cu+Oil+Sediment) provide the clearest signal separation, achieving top F1 scores ($\geq$ 0.88). The model maintains an overall macro-averaged accuracy and F1 of 0.80 across all tertiary mixtures, showing resilience even under complex, three-way chemical interactions.

\begin{table}[ht]
\centering
\scriptsize 
\caption{Classification performance on tertiary mixtures}
\begin{tabular}{|c|c|c|c|c|}
\hline
\textbf{Mixture} & \textbf{Accuracy} & \textbf{Precision} & \textbf{Recall} & \textbf{F1-score} \\
\hline
Cu + Fe + Mg       & 0.70 & 0.75 & 0.72 & 0.73 \\
Cu + Fe + Oil      & 0.80 & 0.82 & 0.78 & 0.80 \\
Cu + Fe + Sediment & 0.78 & 0.80 & 0.76 & 0.78 \\
Cu + Mg + Oil      & 0.79 & 0.81 & 0.77 & 0.79 \\
Cu + Mg + Sediment & 0.77 & 0.79 & 0.75 & 0.77 \\
Cu + Oil + Sediment & 0.88 & 0.90 & 0.87 & 0.88 \\
Fe + Mg + Oil      & 0.76 & 0.78 & 0.74 & 0.76 \\
Fe + Mg + Sediment & 0.74 & 0.76 & 0.73 & 0.75 \\
Fe + Oil + Sediment & 0.85 & 0.87 & 0.84 & 0.85 \\
Mg + Oil + Sediment & 0.84 & 0.86 & 0.83 & 0.84 \\
\hline
\end{tabular}
\label{tab:ternary_mixture_metrics}
\end{table}

These classification results validate the effectiveness of our modeling pipeline, where the phase–AoA tensor captures subtle spatiotemporal variations, and PARAFAC decomposition distills meaningful fingerprints. The student network successfully learns to associate these fingerprints with pollutant presence, even in challenging multi-component mixtures, achieving high precision and recall across a chemically diverse dataset.

\subsection{Regression Performance}
Following the classification evaluation, we now assess the student network’s ability to quantitatively predict the proportion of each pollutant present in a mixture. Table~\ref{tab:rmse_results} reports the mean root-mean-square error (RMSE) between the predicted concentrations $\hat{c}_j$ and ground-truth values $c_j^*$, averaged over present components across all sample types.

\begin{table}[!ht]
\scriptsize
\centering
\begin{minipage}{0.48\linewidth}
\centering
\caption{Prediction performance of student network}
\begin{tabular}{|c|c|}
\hline
\textbf{Sample Type} & \textbf{Mean RMSE} \\
\hline
Pure      & 0.1567 \\
Binary    & 0.2126 \\
Tertiary  & 0.2431 \\
\hline
\end{tabular}
\label{tab:rmse_results}
\end{minipage}
\hfill
\begin{minipage}{0.48\linewidth}
\centering
\caption{Ablation study of feature contributions}
\begin{tabular}{|c|c|c|}
\hline
\textbf{Feature Removed} & \textbf{RMSE} & \textbf{Accuracy} \\
\hline
None (Full Model)        & \textbf{0.2041} & \textbf{0.85} \\
AoA                      & 0.2350          & 0.82 \\
Power                    & 0.2220          & 0.83 \\
Phase                    & 0.5005          & 0.45 \\
\hline
\end{tabular}
\label{tab:ablation}
\end{minipage}
\end{table}


In single-component samples, the RMSE is bounded by residual noise and the tolerance of the Huber loss. Binary mixtures yield slightly higher errors due to partial overlap in component signatures, especially when one constituent is present in lower proportion. Ternary mixtures are more challenging, as multiple overlapping phase--AoA patterns introduce greater ambiguity. Nevertheless, our model maintains a mean per-component RMSE below 0.25, which is comparable to normalized RMSE ranges achieved by standard lab-grade techniques such as Liquid Chromatography–High Resolution Mass Spectrometry (LC-HRMS) ($\approx$0.09--0.25, reported in recent work that benchmarks machine learning models against LC-HRMS outputs~\cite{sepman2023bypassing}). This confirms that \ourmethod offers a competitive level of quantification performance in a fully non-invasive, portable radar-based system.

\subsection{System Evaluation and Benchmarking}
We now present a comprehensive evaluation of \ourmethod\ to validate its performance in pollutant classification, concentration regression, and system robustness. We begin with a comparison against the closest baseline~\cite{salami2023water}, followed by a series of ablation studies that isolate the contribution of key signal features and architectural components. To further demonstrate the resilience of our design, we conduct perturbation experiments involving speaker position, radar tilt, container material, and reflector variation. These evaluations highlight the effectiveness and generalizability of our acoustic–mmWave sensing pipeline across diverse real-world deployment scenarios.

\subsubsection{Comparison with Baseline Models}
\begin{figure}[!ht]
\centering
\includegraphics[width=0.32\textwidth]{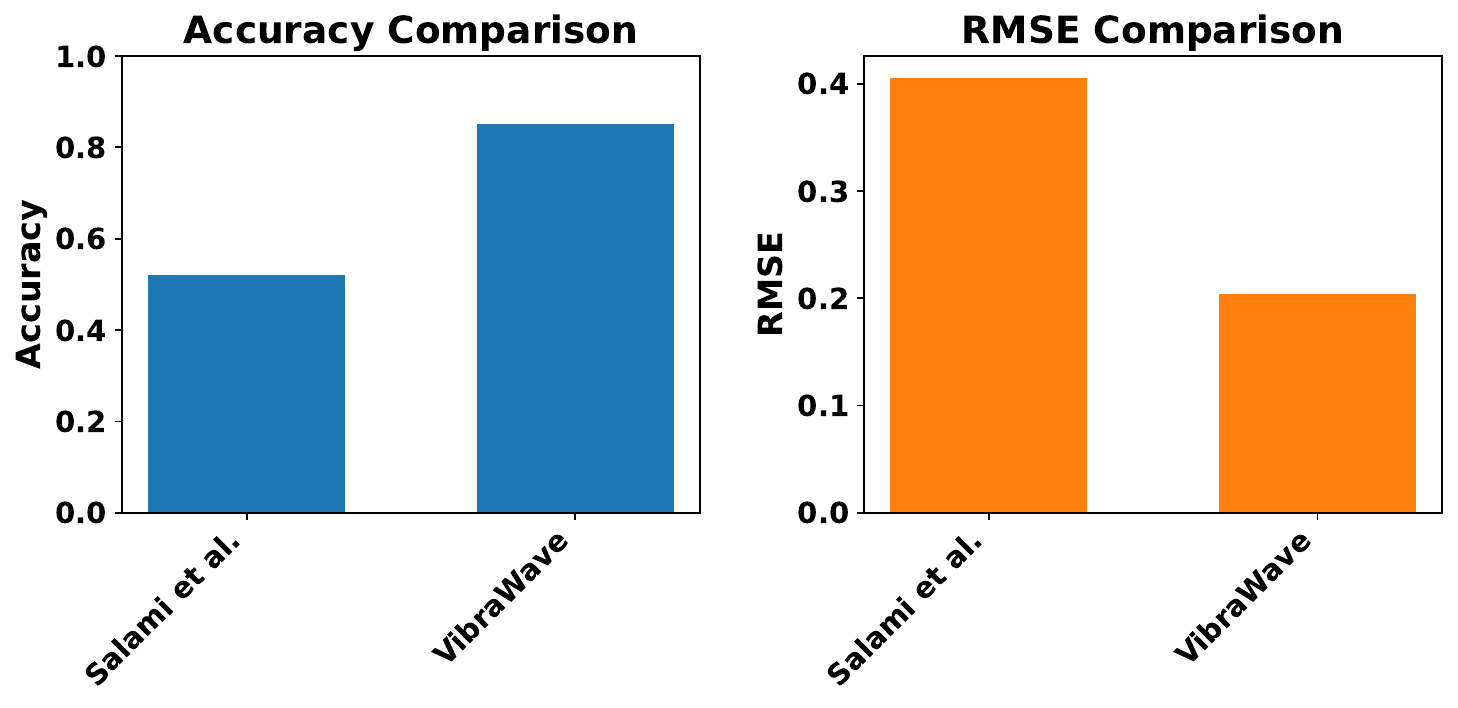}
\caption{Comparison with prior 3D-CNN baseline}
\Description{Diagram showing the comparison with baseline.}
\label{fig:baseline}
\end{figure}

We benchmarked \ourmethod against the closest prior work by Salami et al.~\cite{salami2023water}, which applied a 3D CNN on raw I/Q data for water type classification. We retrained their 3D CNN architecture on our mixture classification task. As shown in Figure~\ref{fig:baseline}, our distilled ResMLP achieves better accuracy and lower RMSE (0.2041), underscoring the benefit of physics-informed tensor decomposition and fingerprinting over raw data-driven approaches.

\subsubsection{Feature Importance via Ablation Study}
To understand the role of each feature group, we ablated angle (AoA), power, and phase features individually. Table~\ref{tab:ablation} shows that removing phase features causes the steepest drop in both RMSE and classification accuracy, confirming their dominant role in encoding pollutant-specific dynamics.


\subsubsection{Architectural Ablations}
We evaluated the contribution of each architectural and training choice in our distillation framework (Table~\ref{tab:ablation-architecture}). Removing KL divergence, Huber loss, or soft labels led to a noticeable increase in RMSE and drop in accuracy, demonstrating the benefit of teacher guidance and robust loss functions. Reducing network depth or removing residuals also degraded performance, confirming the design necessity.

\begin{table}[!ht]
\centering
\caption{Ablation study on architectural components}
\scriptsize 
\begin{tabular}{|c|p{4cm}|c|c|}
\hline
\textbf{Variant} & \textbf{Description} & \textbf{RMSE} & \textbf{Accuracy} \\
\hline
Full (Ours) & KL + Huber + soft-labels + residual + dual-head & \textbf{0.2041} & \textbf{0.85} \\
No KL & Remove KL loss; use only Huber regression & 0.2342 & 0.82 \\
No Huber & Use MSE instead of Huber (with KL) & 0.2287 & 0.83 \\
No soft-labels & Supervised only on hard targets; no teacher guidance & 0.2475 & 0.80 \\
No residual & Remove skip connections & 0.2261 & 0.82 \\
Shallow MLP & Reduce network to 2 layers (128$\rightarrow$64) & 0.2398 & 0.81 \\
\hline
\end{tabular}
\label{tab:ablation-architecture}
\end{table}

\subsubsection{Teacher-Student Model Selection}
\begin{figure}[!ht]
\centering
\includegraphics[width=0.42\textwidth]{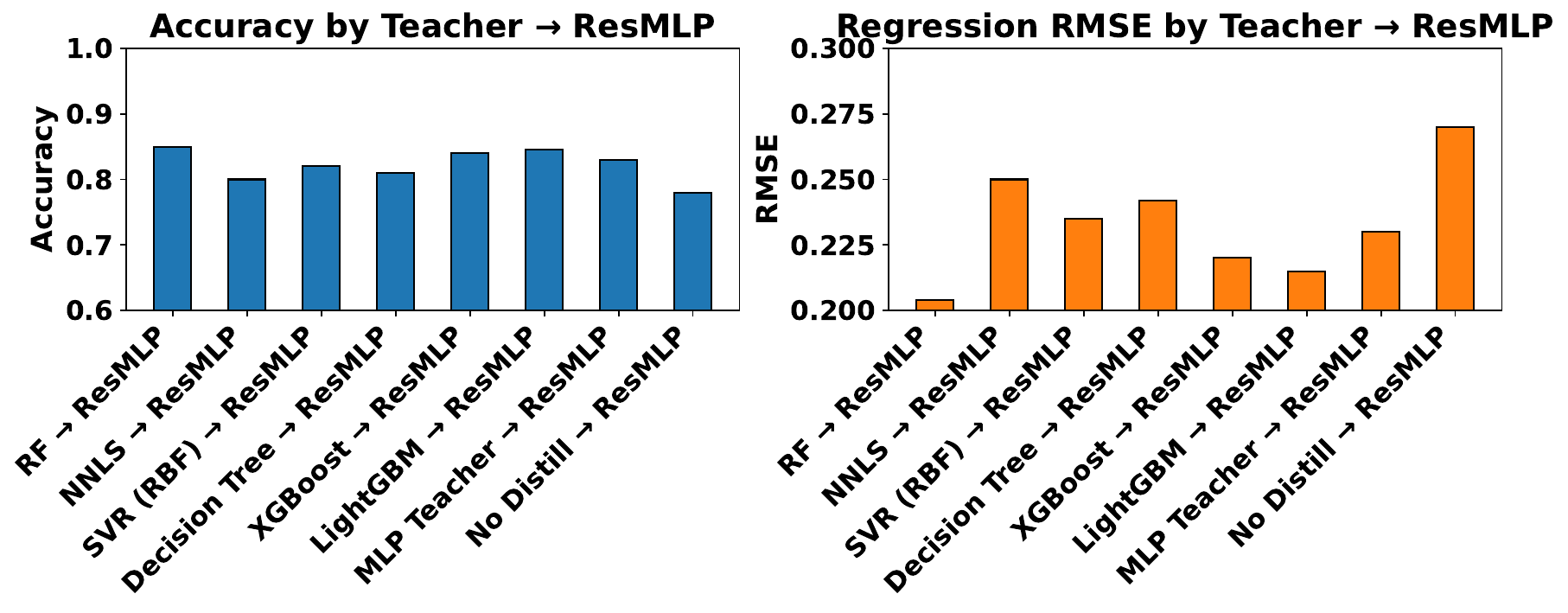}
\caption{Teacher model comparison}
\Description{Diagram showing the teacher model comparison.}
\label{fig:baseline2}
\end{figure}

\begin{figure}[!ht]
\centering
\includegraphics[width=0.42\textwidth]{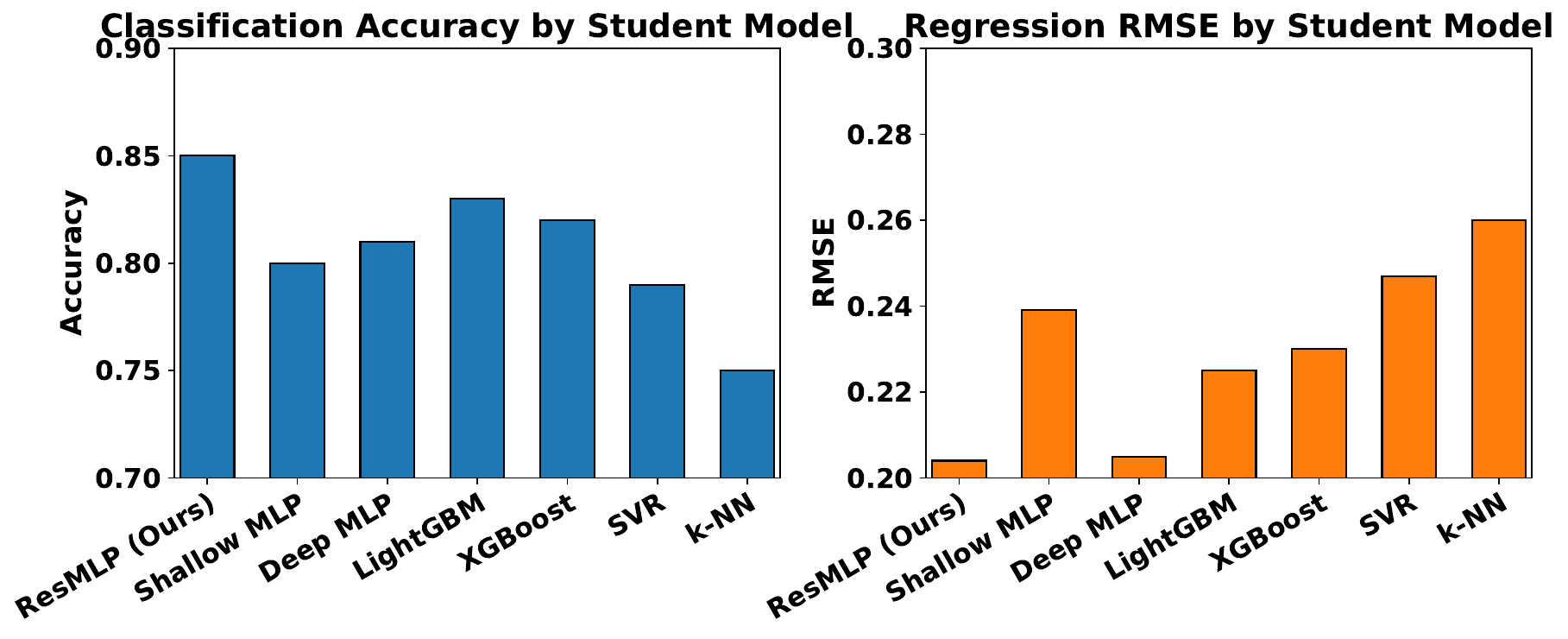}
\caption{Student model comparison}
\Description{Diagram showing the student model comparison.}
\label{fig:baseline3}
\end{figure}

We tested multiple candidate models for both teacher and student networks. Figure~\ref{fig:baseline2} shows that Random Forest provided the best supervision among candidate teachers. Figure~\ref{fig:baseline3} confirms ResMLP as the most accurate and stable student, outperforming CNN and vanilla MLP alternatives.

\subsubsection{Robustness to Radar Misalignment}
In practical deployments, it is not always possible to maintain perfect orthogonal alignment between the radar and the liquid surface. To evaluate the robustness of \ourmethod under such real-world deviations, we intentionally tilted the radar by $\pm$15$^\circ$ relative to the container axis and re-ran inference on the test samples. As shown in Figure~\ref{fig:sidebyside_diagrams}(a), we observed only a minor drop in classification accuracy -- approximately 0.02 for $-15^\circ$ and 0.03 for $+15^\circ$. This minimal degradation suggests that the learned phase–AoA fingerprints remain stable under modest angular perturbations, confirming the angular tolerance of our system and supporting its feasibility in field conditions where precise alignment may not be guaranteed.

\subsubsection{Robustness to Acoustic Source Placement}
\begin{figure}[t]
\centering
\begin{minipage}[t]{0.48\linewidth}
    \centering
    \includegraphics[width=0.8\linewidth]{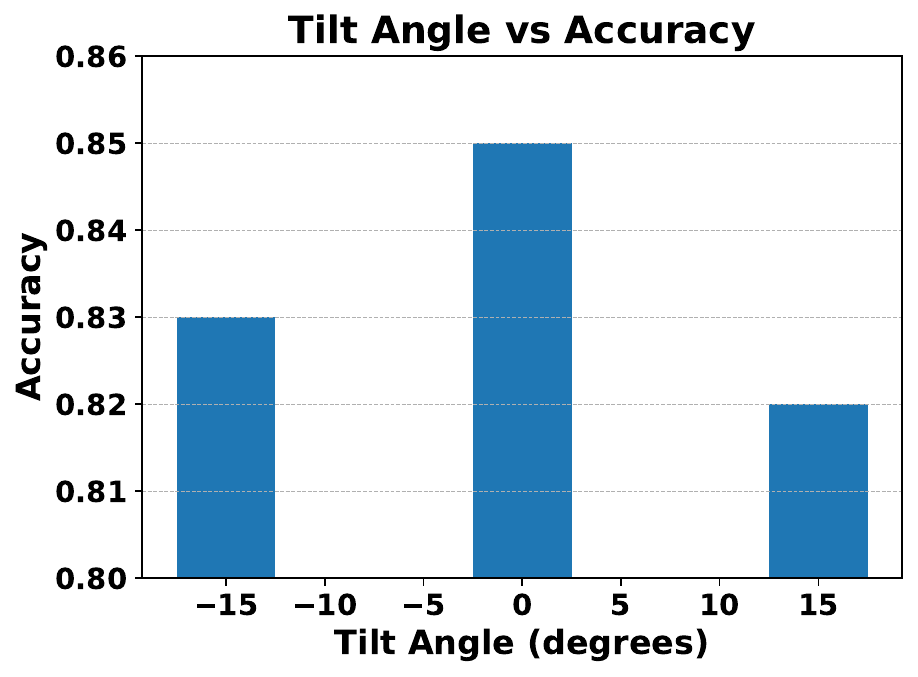}
    \caption*{(a) Radar angle variation}
\end{minipage}
\hfill
\begin{minipage}[t]{0.48\linewidth}
    \centering
    \includegraphics[width=0.9\linewidth]{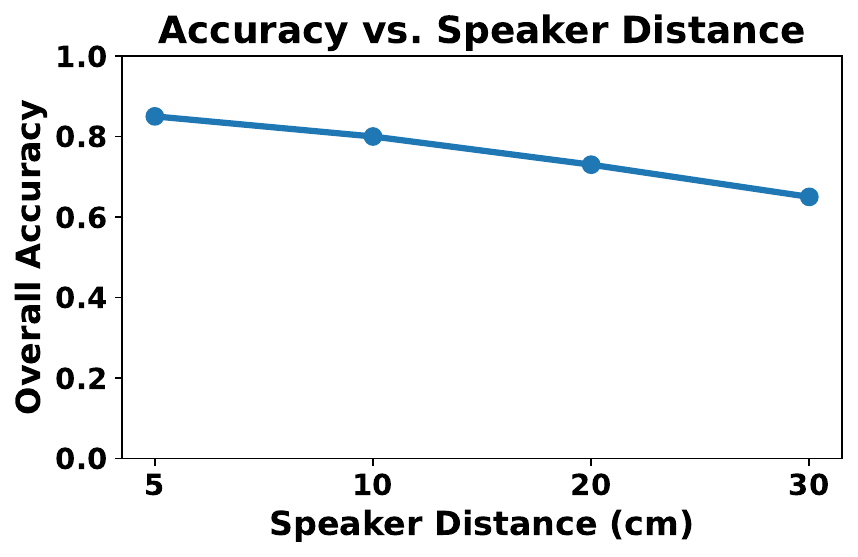}
    \caption*{(b) Accuracy vs. distance}
\end{minipage}
\vspace{-0.5em}
\caption{Angular and spatial perturbations}
\Description{Diagram showing the speaker accuracy and radar angle variation.}
\label{fig:sidebyside_diagrams}
\end{figure}

The effectiveness of our approach hinges on the ability of acoustic vibrations to excite measurable phase shifts on the liquid surface. To understand how the placement of the speaker affects this mechanism, we varied the distance between the acoustic source and the water surface from 5~cm to 30~cm. As shown in Figure~\ref{fig:sidebyside_diagrams}(b), classification accuracy degrades gradually—from 0.85 at 5~cm to 0.80 at 10~cm, and further down to 0.65 at 30~cm. This trend is attributed to weaker acoustic coupling at larger distances, which reduces the strength of induced surface displacements and thus the observable phase modulation in radar returns. Our results indicate that for optimal performance, the speaker should be positioned within 10~cm of the liquid interface to ensure sufficient vibration energy and reliable sensing.

\subsubsection{Speaker Power Effects}
The strength of the acoustic excitation plays a critical role in the effectiveness of our sensing pipeline, as the vibrational energy induced in the liquid directly affects the radar-measured phase modulations. To evaluate this, we compared two commercially available Bluetooth speakers with distinct power ratings: the JBL Flip SE 3 (16~W) and the Sony SRS-XB100 (5~W). The JBL Flip SE 3 features dual passive radiators and a robust driver setup designed for high-output audio, while the Sony model is a compact, portable speaker optimized for energy efficiency and casual listening.

\begin{figure}[ht]
\centering
\includegraphics[width=0.35\textwidth]{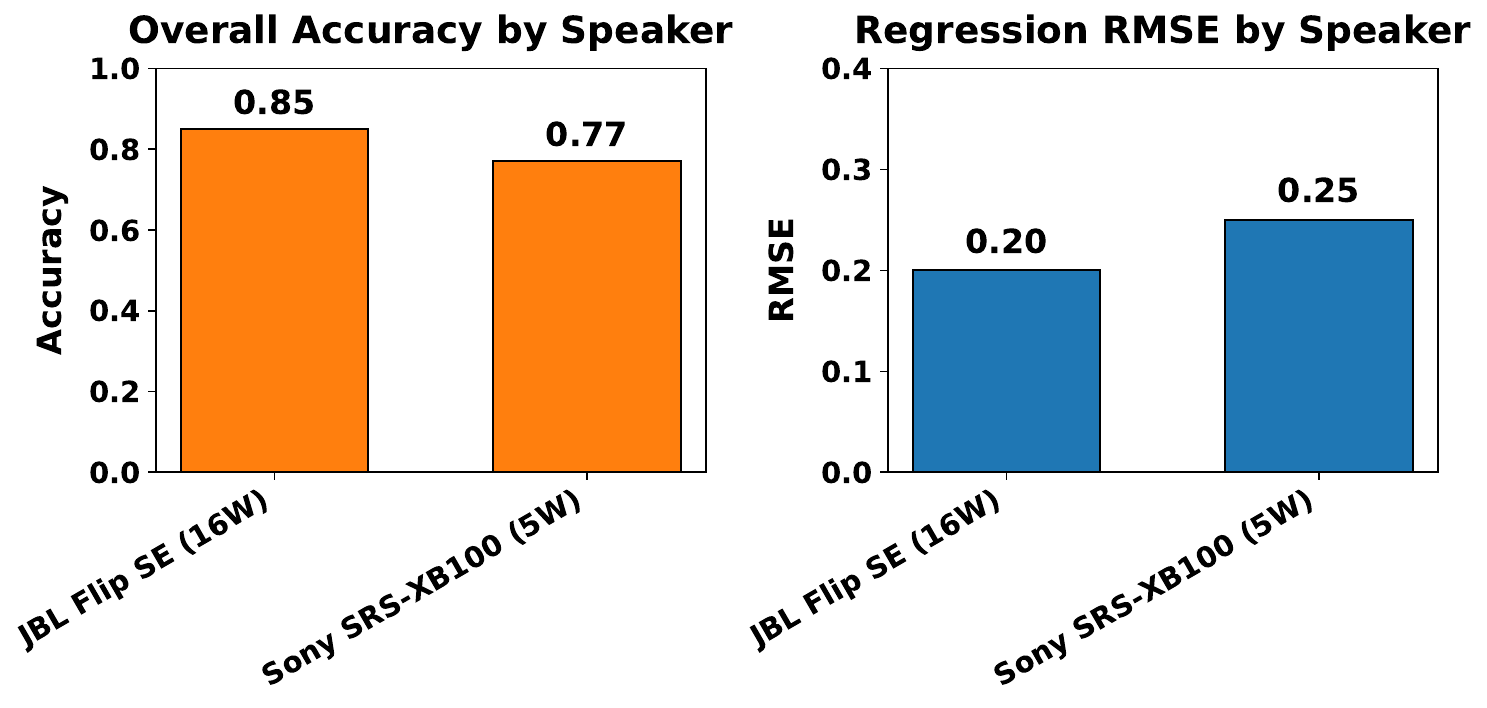}
\caption{Performance with different speakers}
\label{fig:speaker}
\end{figure}

As shown in Figure~\ref{fig:speaker}, the higher-power JBL speaker yielded noticeably better performance, achieving a classification accuracy of 0.85 and RMSE of 0.20. In contrast, the Sony speaker, due to its lower acoustic output, resulted in a reduced accuracy of 0.77 and a higher RMSE of 0.25. This performance drop stems from weaker acoustic coupling with the water surface, which leads to smaller surface displacements and consequently, less pronounced phase changes in the radar returns. These results underscore the importance of using a sufficiently powerful acoustic source to ensure reliable vibrational excitation, especially in settings with subtle or low-amplitude pollutant-induced responses.

\subsubsection{Variation Across Reflectors}
To assess the sensitivity of our system to changes in the underlying reflective surface, we evaluated performance across three reflector materials commonly found in real-world settings: a book (paper-based), a metal sheet (aluminum), and a wood panel. These materials vary significantly in electromagnetic reflectivity and surface texture -- metal offers high reflectivity and smoothness, wood is more absorptive and irregular, while a book lies somewhere in between.

\begin{figure}[ht]
\centering
\includegraphics[width=0.32\textwidth]{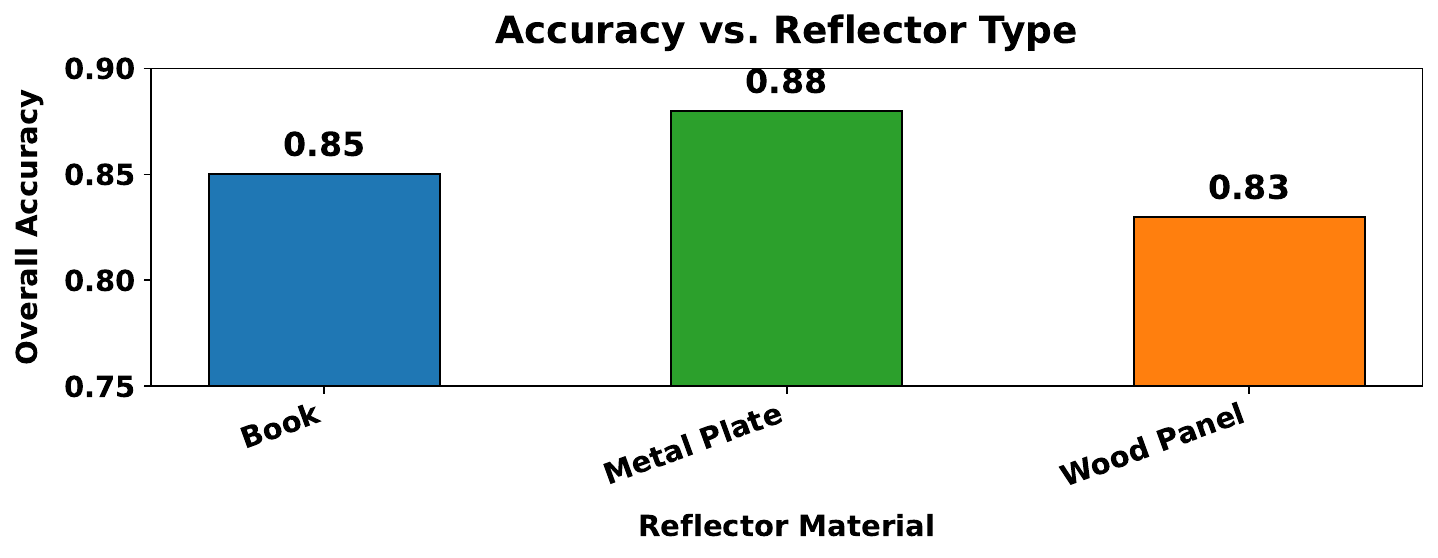}
\caption{Performance across reflector materials}
\Description{Diagram showing the peformance across reflectors.}
\label{fig:Accuracy Variation across reflectors}
\end{figure}

As shown in Figure~\ref{fig:Accuracy Variation across reflectors}, classification accuracy varied slightly—0.85 with the book (our default baseline), 0.84 with the metal reflector, and 0.83 with wood. This small $\pm$0.01 to $\pm$0.03 variation confirms that the \ourmethod pipeline is robust to differences in reflector material. Despite changes in reflected signal strength or multi-path characteristics due to surface roughness and dielectric properties, the learned phase–AoA fingerprints remain distinctive and resilient. This adaptability ensures reliable pollutant detection even when the system is deployed on non-ideal or uncalibrated surfaces.

\subsubsection{Impact of PARAFAC Unmixing}
Table~\ref{tab:unmixing_comparison} presents a comparative evaluation of different unmixing strategies, highlighting the role of PARAFAC in enabling effective signal decomposition. When only raw radar features are used without dimensionality reduction or decomposition, both classification accuracy and regression performance degrade substantially due to noise and mixed-source entanglement in the data. Applying non-negative least squares (NNLS) directly on raw features yields modest improvements, but remains limited in interpretability. Incorporating PCA for dimensionality reduction before NNLS performs slightly better, but still lacks component specificity. In contrast, PARAFAC decomposition followed by NNLS achieves the highest accuracy (0.85) and the lowest RMSE (0.20), clearly indicating that the multilinear structure of the radar response, captured across frequency tones, range bins, and spatial angles, is best represented in tensor form. This confirms that tensor factorization is not only mathematically principled but practically essential for isolating latent pollutant-specific fingerprints in complex mixtures.

\begin{table}[ht]
\centering
\caption{Comparison of pollutant unmixing methods}
\scriptsize
\begin{tabular}{|l|c|c|}
\hline
\textbf{Method} & \textbf{Accuracy} & \textbf{RMSE} \\
\hline
Raw data                 & 0.55 & 0.47 \\
Raw + NNLS               & 0.68 & 0.35 \\
PCA (10 PCs) + NNLS      & 0.74 & 0.29 \\
PARAFAC + NNLS (Ours)    & \textbf{0.85} & \textbf{0.20} \\
\hline
\end{tabular}
\label{tab:unmixing_comparison}
\end{table}

\subsubsection{Container Material Effects}
We evaluated model performance using two alternative containers of the glass container used in previous experiments: a melamine bowl and a paper cup. As shown in Table~\ref{tab:container_comparison}, both accuracy and RMSE degraded compared to the glass baseline. The melamine bowl’s wide opening and highly reflective surface caused multipath interference; the paper cup's thin walls and higher EM absorption led to signal loss. Nonetheless, the full pipeline remained functional with modest accuracy drops.

\begin{table}[ht]
    \caption{Accuracy and RMSE across container types}
    \centering
    \scriptsize 
    \begin{tabular}{|c|cc|cc|cc|}
        \hline
        \multirow{2}{*}{\textbf{Variant}} &
        \multicolumn{2}{c|}{\textbf{Glass}} &
        \multicolumn{2}{c|}{\textbf{Paper Cup}} &
        \multicolumn{2}{c|}{\textbf{Melamine Bowl}} \\
        \cline{2-7}
        & \textbf{Accuracy} & \textbf{RMSE} & \textbf{Accuracy} & \textbf{RMSE} & \textbf{Accuracy} & \textbf{RMSE} \\
        \hline
        Full Model & 0.85 & 0.20 & 0.78 & 0.30 & 0.68 & 0.48 \\
        MSE + BCE Loss & 0.76 & 0.34 & 0.70 & 0.39 & 0.57 & 0.53 \\
        Without PARAFAC & 0.68 & 0.35 & 0.64 & 0.44 & 0.49 & 0.57 \\
        \hline
    \end{tabular}
    \label{tab:container_comparison}
\end{table}

\section{Discussion}

\noindent
In this section, we critically reflect on the limitations of \ourmethod and identify several edge cases that may be of interest to the systems community. We also outline potential solutions and research directions to address these challenges.

$\bullet\;\;$\noindent{\textbf{Container Dependency and Material Interference:}}
While our system performs well across common container types (e.g., glass, melamine, paper), the radar response is influenced by the geometry and dielectric properties of the container. Materials with high absorbency or irregular surfaces (e.g., thick ceramics or porous plastics) may distort the acoustic field and radar reflections. To mitigate this, future work could explore container-invariant normalization using reference responses or synthetic data augmentation. Domain adaptation and adversarial training may further enhance robustness across diverse vessel types.

$\bullet\;\;$\noindent{\textbf{Sensitivity to Minor Component Concentrations:}}
The model's regression performance degrades when estimating the presence of low-concentration secondary or tertiary pollutants. This is due to dominant signal features from the major component overshadowing subtler modulations induced by trace constituents. Contrastive learning or attention-based mechanisms may help highlight weaker contributions. Additionally, multi-pass radar capture or spectral amplification in post-processing could improve sensitivity to minor concentrations.

$\bullet\;\;$\noindent{\textbf{Acoustic Coupling Variability:}}
The strength and consistency of acoustic-induced phase signatures depend on effective coupling between the speaker and the fluid medium. We observed significant performance drops when using low-power speakers or imprecise placement. Embedding a piezoelectric actuator beneath the container could offer more controlled and reproducible excitation. Alternatively, integrating dynamic gain control based on measured vibrational response can help adaptively adjust radar integration to maintain signal quality.

$\bullet\;\;$\noindent{\textbf{Limitations in Multi-Pollutant Resolution:}}
Our PARAFAC + NNLS pipeline assumes linear mixing in the radar signal space. However, in real-world scenarios, pollutants may interact chemically or physically in non-linear ways (e.g., emulsification, ion pairing). Such interactions could distort the additive model and hinder accurate unmixing. Future directions include kernel-based nonlinear unmixing or physics-guided learning architectures to better handle non-additive pollutant behaviors. Empirical modeling with reactive mixtures could also help capture these effects.

$\bullet\;\;$\noindent{\textbf{Generalization to Unseen Pollutants:}}
The current model is trained on a finite set of pollutants. While it generalizes well to binary and ternary combinations within this set, it lacks mechanisms for detecting unknown contaminants. Zero-shot or few-shot learning paradigms, along with anomaly detection using autoencoder residuals, offer potential solutions. Expanding the training set with surrogate pollutants or generating synthetic phase signatures could further strengthen model ability to handle unseen compositions.

$\bullet\;\;$\noindent{\textbf{Environmental and Operational Robustness:}}
While we validated the system under controlled indoor conditions (temperature 18--36\degree C), field deployments may introduce environmental noise, such as ambient vibration, humidity, or unstable power sources, that could degrade signal fidelity. Embedding inertial sensors for environmental feedback and implementing periodic calibration routines may improve robustness. Additionally, physical stabilization using enclosures or damping mounts could minimize external perturbations.

$\bullet\;\;$\noindent{\textbf{Scalability and Real-Time Operation:}}
Although our distilled ResMLP model is lightweight and achieves fast inference, real-time deployment on constrained edge devices may be bottlenecked by the PARAFAC decomposition step. A practical solution is to shift tensor decomposition to the training phase, precomputing pollutant dictionaries offline. During inference, only a forward pass through the student network would be required, enabling low-latency operation on microcontrollers or embedded systems.

$\bullet\;\;$\noindent\textbf{Limitations and Field Considerations:}
While our current evaluation is limited to controlled pollutant concentrations in distilled water, this setup enables precise attribution of radar signatures to specific contaminants. Extending this approach to field samples (e.g., tap or river water) is a natural next step and may require calibration or domain adaptation techniques, which we plan to explore in future deployments.

\section{Conclusion}
This paper presents \ourmethod, a novel mmWave radar-based framework for non-invasive, real-time monitoring of water pollutants using controlled acoustic excitation, tensor decomposition, and knowledge-distilled deep learning. By modeling the complex interactions between low-frequency vibrations and electromagnetic scattering from containerized liquids, we extract rich phase and AoA features that encode pollutant-specific signatures. Our pipeline, built upon PARAFAC-based tensor factorization and non-negative unmixing, enables interpretable fingerprinting and accurate quantification of multiple co-occurring contaminants. Extensive experiments across diverse mixtures and environmental conditions demonstrate that \ourmethod achieves classification accuracy of 0.85 and per-component regression RMSE of 0.20, comparable to laboratory-grade analytical techniques. With its compact implementation, data efficiency, and robust generalization, \ourmethod offers a promising step toward scalable, field-deployable water quality sensing using commodity radar platforms. Our method is pollutant-agnostic in design, relying on material-induced variations in dielectric response and surface resonance. Extending the model to broader pollutant classes or mixtures will require additional training data and possibly transfer learning techniques, which we identify as promising directions for future work.


\newpage
\bibliographystyle{ACM-Reference-Format}
\bibliography{mybib}
\end{document}